\documentclass[12pt]{article}

\usepackage{cmap}
\usepackage[T2A]{fontenc}
\usepackage[utf8]{inputenc}
\usepackage[english, ngerman]{babel}
\usepackage{float}
\usepackage{afterpage}

\usepackage{amsfonts}
\usepackage{amsthm}
\usepackage{amstext}
\usepackage{amsmath}
\usepackage{amssymb}
\usepackage{booktabs}
\usepackage{indentfirst} 
\usepackage{enumerate}
\usepackage[font=small,skip=0pt]{caption}

\usepackage[numbers,square,sort]{natbib} 
\bibpunct{(}{)}{,}{a}{,}{,} 

\usepackage{graphics}
\usepackage[pdftex]{graphicx}
\usepackage{color}
\usepackage{lscape} 
\usepackage{bm} 
\usepackage{tikz}
\usepackage{titlesec} 
\usepackage{float} 
\usepackage{longtable}
\usepackage{booktabs}
\usepackage[center]{subfigure}
\usepackage{captcont}
\usepackage{tabularx}
\usepackage{diagbox} 
\usepackage{rotating}
\usepackage{geometry}
\usepackage[super]{nth}
\usepackage{comment} 
\usepackage{textcomp}

\theoremstyle{definition}


\renewcommand\appendix{\par
  \setcounter{section}{0}
  \setcounter{subsection}{0}
  \setcounter{figure}{0}
  \setcounter{table}{0}
  \renewcommand\thesection{Appendix \Alph{section}}
  \renewcommand\thefigure{\Alph{section}\arabic{figure}}
  \renewcommand\thetable{\Alph{section}\arabic{table}}

}

\begin{document}
\title{\textbf{COVID-19: Tail Risk and Predictive Regressions}\footnote{The authors are grateful to Paul Kattuman, Ulrich K. M{\" u}ller and the participants at the the 2021 International Association for Applied Econometrics (IAAE-2021) Annual Conference, the Joint Meeting of the 2-nd Workshop in Applied Econometrics and the VII-th International Conference on Modern Econometrics Tools and Applications (META2020) and the Centre for Business Analysis (CEBA, St. Petersburg State University) seminar series for helpful comments and discussions. We also thank Ronald Huisman for kindly sharing the SciLab codes for implementation of weighted Hill's tail index estimates in \cite{huisman2001tail}. Rustam Ibragimov and Anton Skrobotov gratefully acknowledge the support provided by the Russian Foundation for Basic Research, Project No. 20-010-00960.}}

\author{{\Large{Walter Distaso}} \\Imperial College London\\Business School  \and {\Large{Rustam Ibragimov}} \\Imperial College London \\Business School \and {\Large{Alexander Semenov}} \\St. Petersburg State University and \\ Herbert Wertheim  College of Engineering, the University of Florida \and {\Large{Anton Skrobotov}} \\ Russian Presidential Academy of National Economy \\and Public Administration and St. Petersburg State University}
\date{\empty} \maketitle \begin{abstract}  \noindent
The paper focuses on econometrically justified robust analysis of the effects of the COVID-19 pandemic on financial markets in different countries across the World. It provides the results of robust estimation and inference on predictive regressions for returns on major stock indexes in 23 countries in North and South America, Europe, and Asia incorporating the time series of reported infections and deaths from COVID-19. We also present a detailed study of persistence, heavy-tailedness and tail risk properties of the time series of the COVID-19 infections and death rates that motivate the necessity in applications of robust inference methods in the analysis. Econometrically justified  analysis is based on heteroskedasticity and autocorrelation consistent (HAC) inference methods, 
	recently developed robust $t$-statistic inference approaches and robust tail index estimation.
\end{abstract}

\maketitle

\newpage

\section{Introduction}
Several recent papers have focused on econometric and statistical analysis and forecasting of key time series and variables associated with the on-going COVID-19 pandemics, including infection and death rates, and their effects on economic and financial markets (see \citeauthor{Stock1}, \citeyear{Stock1}, \citeauthor{Toda1}, \citeyear{Toda1}, \citeauthor{Miles}, \citeyear{Miles}, \citeauthor{Villaverde}, \citeyear{Villaverde}, \citeauthor{Paul},  \citeyear{Paul},  \citeauthor{Pesaran}, \citeyear{Pesaran}, \citeauthor{Stock}, \citeyear{Stock},   \citeauthor{Toda}, \citeyear{Toda}, \citeauthor{Linton}, \citeyear{Linton},  \citeauthor{Manski}, \citeyear{Manski}, among others).

This paper contributes to the above literature by focusing on the robust analysis of the effects of the pandemics on financial markets across the World. We provide the results of robust estimation and inference on predictive regressions for returns on major stock indexes in 23 developed and emerging economies in North and South America, Europe, and Asia incorporating the time series of reported infections and deaths from COVID-19.

We also present a detailed study of persistence, heavy-tailedness and tail risk properties of COVID-19 infections and deaths time series that emphasize the necessity in applications of robust inference methods in the analysis and forecasting of the COVID-19 pandemic and its impact on economic and financial markets and the society.

Econometrically justified and robust analysis in the paper is based on  heteroskedasticity and autocorrelation consistent (HAC) inference methods, 
recently developed robust $t$-statistic inference procedures and robust tail index estimation approaches.

The results of the analysis, in particular, point to potential non-stationarity in the form of unit roots in the time series of daily infections and deaths from COVID-19 that are commonly used in research on modelling and forecasting of the COVID-19 pandemic and its effects. The results emphasize the necessity in basing the analysis of models incorporating the COVID-19-related time series such as daily infection and death rates on their (stationary) differences. The analysis using robust tail index inference methods further indicates potential heavy-tailedness with possibly infinite variances and first moments in the time series of daily infections and deaths from COVID-19 and their differences in countries across the World.

In order to account for the problems of potential non-stationarity in the daily COVID-19 infections and deaths time series, the paper provides the analysis of predictive regressions for financial returns incorporating both the lagged daily infections/deaths from COVID-19 and their differences. Further, the properties of autocorrelation, heavy-tailedness and heterogeneity in the COVID-19 infections/deaths time series are accounted for by the use in the predictive regression analysis of both the widely applied standard HAC inference methods as well as the recently proposed $t-$statistic approaches to robust inference under the above problems in the data (see \citeauthor{IM}, \citeyear{IM, IM1}, Section 3.3 in \citeauthor{ibragimov2015heavy}, \citeyear{ibragimov2015heavy} and Section \ref{pred1} in this paper).

The standard HAC inference methods indicate (apparently spurious) statistical significance of the (potentially non-stationary) lagged daily infections and deaths from COVID-19 in predictive regressions for returns on the major stock indices in some countries. HAC methods also point to  statistical significance of (stationary) lagged daily changes in the number of COVID-19 infections and deaths for some countries. For daily changes in COVID-19 infections, high statistical significance, with the expected negative signs of the predictive regression coefficients, is observed in the case of Italy, India, Brazil and Argentina.

Motivated by the results on high persistence and heavy-tailedness in daily COVID-19 infections/ deaths time series obtained in the paper and also by poor finite sample properties of HAC inference methods (see Section \ref{pred1} and references therein), we further provide the assessment of statistical significance of the coefficients in the predictive regressions using $t-$statistic approaches to robust inference based on group estimates. According to the statistically justified analysis using the robust $t-$statistic approaches, the lagged daily COVID-19 infection and death rates and their (stationary) differences  appear not to be statistically significant in predictive regressions for stock index returns in all the countries considered in the analysis.

Overall, one of the main conclusions from the results in the paper is that statistical and econometric analyses and forecasts of the on-going COVID-19 pandemic and its impacts on economic and financial markets and the society should be based on theoretically justified robust inference methods. The methods used in the analysis and the forecasting of the pandemic and its effects should account, in particular, for the problems of potential non-stationarity, autocorrelation, heavy-tailedness and heterogeneity in the key time series and variables related to COVID-19, including the infections and deaths time series.



The paper is organised as follows. Section \ref{UR} provides the results of (non-)stationarity and unit root tests for the time series of COVID-19 infection and death rates in the countries across the World.  Section \ref{data} describes the data used in the analysis. Section \ref{heavy} presents the analysis of heavy-tailedness and tail risk properties of daily COVID-19 infection and death rates. Section \ref{pred1} provides the results of theoretically justified and robust statistical analysis of predictive regressions for the returns on major stock indices in the countries considered incorporating the time series on COVID-19 infections and deaths. Section \ref{conclude} makes some concluding remarks and discusses directions for further research. Appendices A and B provide the diagrams and tables on the results of the statistical analysis in the paper.

\section{Data} \label{data}

The analysis in the paper uses the data on daily COVID-19 infections and deaths in different countries across the World (the UK, Germany, France, Italy, Spain, Russia, the Netherlands, Sweden, India, Austria, Finland, Ireland, the US, Lithuania, Canada, Brazil, Mexico, Argentina, Japan, China, South Korea, Indonesia and Australia) for the period from 22 January 2020 to 22 March 2021. The data is obtained from the Data Repository maintained by the Center for Systems Science and Engineering (CSSE) at Johns Hopkins University.\footnote{https://github.com/CSSEGISandData/COVID-19} The data on daily prices of major stock indices for the countries considered is obtained from Yahoo Finance and the data on interest rates is from the Global Rates database.\footnote{https://www.global-rates.com/interest-rates/central-banks/central-banks.aspx}
We consider the following stock indices: FTSE 100 (UK), DAX (Germany), CAC 40 (France), FTSE MIB (Italy), IBEX 35 (Spain), MOEX (Russia), AEX (Netherlands), OMXS 30 (Sweden), SENSEX (India), ATX (Austria), OMX Helsinki 25 (Finland), ISEQ (Ireland), Dow Jones, S\&P 500 (USA), OMX Vilnius (Lithuania), TSX (Canada), iBovespa (Brazil), IPC Mexico (Mexico), Merval (Argentina), NIKKEI 225 (Japan), SHANGHAI (China), KOSPI (South Korea), JCI (Indonesia), ASX 50, ASX 200 and Australian All Ordinaries (Australia). The analysis uses central bank rates for the countries considered; European interest rate is used for the country members of European monetary union.

Throughout the paper, $I_t$ and $D_t$ denote the (cumulative) number of COVID-19 infections and deaths from the beginning of the period on 22 January 2020 to day $t$ in the countries considered. Further, $\Delta I_t$ and $\Delta D_t$ denote the differences of the above time series, that is the number of reported infections and deaths in day $t.$ By $\Delta^2 I_t$ and $\Delta^2 D_t$ we denote the cumulative infections/deaths time series' second differences, that is, the daily changes in the number of COVID-19 infections and deaths in the countries dealt with. The estimation and testing in the paper is based on the periods with positive values of the number of COVID-19 infections and deaths $I_t$ and $D_t$ in the countries considered.

\section{Empirical results}

\subsection{(Non-)stationarity analysis} \label{UR}
We begin the analysis by the study of the degree of integration in the time series $I_t,$ $D_t,$ $\Delta I_t,$  $\Delta D_t,$ $\Delta^2 I_t$ and $\Delta^2 D_t$ of COVID-19 infections and deaths and their differences in the countries considered. Tables \ref{tab1} and \ref{tab11} present the results of several unit root tests for the time series of daily infections and deaths $\Delta D_t, \Delta D_t$ and the time series of daily changes in their number $\Delta^2 I_t, \Delta^2 D_t$ . The results are provided for the (right tailed) likelihood ratio unit root test proposed by \citet{JanssonNielsen2012} (with the test statistic $LR$ in Tables \ref{tab1} and \ref{tab11}; see also \citeauthor{skrobotov2018bootstrap}, \citeyear{skrobotov2018bootstrap}), the GLS-based modified Phillips-Perron type tests (with the corresponding test statistics $MZ_\alpha$, $MSB$, $MZ_t$), the modified point optimal test (with the test statistic $MP_t;$ see \citeauthor{NP2001}, \citeyear{NP2001}) and the GLS-based Augmented Dickey-Fuller test (with the test-statistic denoted by $ADF$ in Tables \ref{tab1} and \ref{tab11}; see \citeauthor{ERS1996}, \citeyear{ERS1996}). To address the issue of possible heavy tails and infinite variance of the series (see the next section), for calculation of the $p-$value of the unit root tests, we use recently justified sieve wild bootstrap algorithm with a Rademacher distribution employed in the wild bootstrap re-sampling scheme (see \citeauthor{cavaliere2016unit}, \citeyear{cavaliere2016unit}).

An important tuning parameter in the above tests is related to the choice of lag length used in the analysis. We use the modified Akaike information criterion (MAIC) lag choice approach based on standard ADF regressions as suggested by \citet{PQ2007}. According to the results (the wild bootstrap $p-$values are given in brackets), the unit root hypothesis in the time series $\Delta I_t$ of daily COVID-19 infections is not rejected at reasonable significance levels, e.g., 5\% and 10\%, by all the employed tests for all the countries considered except Spain, Sweden, Ireland, China and South Korea. For the time series $\Delta D_t$ of daily COVID-19 related deaths, the unit root hypothesis in  is not rejected for all the countries considered except Spain, Sweden, Finland, Argentina and China. For the daily infections and deaths time series $\Delta I_t$ and $\Delta D_t$ in China, the rejection of the unit root hypothesis is on every reasonable significance level (even at 1\%). 

On the other hand, according to the results in Tables \ref{tab1} and \ref{tab11}, the unit root hypothesis is rejected at all reasonable significance levels by all the tests for the time series $\Delta^2 I_t$ and $\Delta^2 D_t$ of daily changes in the number of COVID-19 infections and deaths in all the countries considered.


The above results of unit root tests have several important implications for statistical analysis of models and key time series related to the COVID-19 pandemic and its effects. According to the results, in the countries across the World, the time series of daily COVID-19 infections and deaths and thus the time series of total (cumulative) infections/deaths from the disease up to a certain date that are typically employed in the analysis and forecasting of the pandemic and its impact appear to exhibit non-stationarity. The daily COVID-19 infections/deaths time series $\Delta I_t$ and $\Delta D_t$ appears to exhibit unit root process persistence for most of the countries considered. This, in turn, implies very high persistence in the time series $I_t$ and $D_t$ of total infections/deaths up to a certain date that appear to be integrated of order 2.\footnote{The conclusions on persistence properties of the time series $I_t, D_t$ and $\Delta I_t, \Delta D_t$ are somewhat similar to those for the CPI and the inflation rate (the change in the logarithm of the CPI) time series, where often unit root hypothesis is not rejected for the inflation rate and thus the (logarithm) of the CPI levels appears to be integrated of order 2 (see the analysis of non-stationarity in Section 14.6 in \citeauthor{SW}, \citeyear{SW}, for the inflation rate and its changes in the US). These conclusions imply the necessity of the use of differences of the inflation rate in time series modeling of inflation and its relationship to other key economic variables such as the unemployment level in the Phillips curve (see Chs. 14 and 16 in \citeauthor{SW}, \citeyear{SW}).}


Non-stationarity of daily COVID-19 infections and deaths time series implies that, due to the spurious regression problem, 
the statistical analysis of models incorporating these and other nonstationary variables related to the pandemic should be based on their stationary differences 
as in the case of predictive regressions for financial returns in Section \ref{pred1}.

\subsection{Heavy-tailedness and tail risk analysis} \label{heavy}
In this section, we provide the analysis of heavy-tailedness and tail risk properties of daily COVID-19 infection and death rates in the countries considered. The estimates point to pronounced heavy-tailedness in the infections/deaths time series. This further motivates the necessity in applications of robust methods in modelling and forecasting the dynamics of infection and death rates and other variables related to the pandemic and inference on their effects on the world financial and economic markets.

As indicated in many empirical and theoretical works in the literature (see, among others, the analysis and the reviews in \citeauthor{embrechts1997modelling},  \citeyear{embrechts1997modelling}, 
\citeauthor{cont2001empirical}, \citeyear{cont2001empirical}, \citeauthor{gabaix2009power}, \citeyear{gabaix2009power}, 
\citeauthor{ibragimov2015heavy}, \citeyear{ibragimov2015heavy}, and references therein), distributions of many variables related to or affected by crises and natural disasters and characterised by the presence of extreme values and outliers, such as financial returns, catastrophe risks or economic losses from natural catastrophes, exhibit deviations from Gaussianity in the form of heavy power law tails. For a positive heavy-tailed variable (e.g., representing a risk, the absolute value of a financial return, or a loss from a natural disaster $X$) one has
\begin{eqnarray} \label{power} P(X>x)\sim \frac {C} {x^{\zeta}} \end{eqnarray}
for large $x>0,$ with a constant $C>0$ and the parameter $\zeta>0$ that is referred to as the tail index (or the tail exponent) of $X.$ The value of the tail index parameter $\zeta$ is important as it characterises the probability mass (heaviness and the rate of decay) in the tails of power law distribution (\ref{power}). Heavy-tailedness (i.e., the tail index $\zeta$) of the variable $X$ governs the likelihood of observing extremes and outliers in the variables. The smaller values of the tail index $\zeta$ correspond to a higher degree of heavy-tailedness in $X$ and, thus, to a higher likelihood of observing extremely large values in realisations of the variable. In addition, importantly, the value of the tail index $\zeta$ governs finiteness of moments of $X,$ with the moment $EX^p$ of order $p>0$ of the variable being finite: $EX^p<\infty$ if and only if $\zeta>p.$ In particular, the variance of $X$ is defined and is finite if and only if $\zeta>2,$ and the first moment of the variable is finite if and only if $\zeta>1.$


The degree of heavy-tailedness and finiteness of variances for variables 
is crucial for applicability of standard statistical and econometric approaches, including regression and least squares methods. Similarly, the problem of potentially infinite fourth moments of (economic and financial) time series dealt with needs to be taken into account in applications of autocorrelation-based methods and related inference procedures in their analysis (see the discussion in \citeauthor{cont2001empirical}, \citeyear{cont2001empirical}, Ch. 1 in \citeauthor{ibragimov2015heavy}, \citeyear{ibragimov2015heavy}, 
and references therein).

Many recent studies argue that the tail indices $\zeta$ in heavy-tailed models (\ref{power}) typically lie in the interval $\zeta\in (2, 4)$ implying finite variances and infinite fourth moments for financial returns in developed economies and maybe smaller than 2 implying possibly infinite variances for financial returns in emerging markets  (see, among others, \citeauthor{LP}, \citeyear{LP},  
\citeauthor{gabaix2009power}, \citeyear{gabaix2009power}, \citeauthor{ibragimov2015heavy}, \citeyear{ibragimov2015heavy}, and references therein).\footnote{Heavy-tailed power law behavior is also exhibited by many other such important economic
and financial variables as income and wealth (with $\zeta\in (1.5, 3)$ and $\zeta\approx 3,$ respectively; see, among others,  \citeauthor{gabaix2009power}, \citeyear{gabaix2009power}, and the references therein); financial returns from technological innovations, losses from operational risks and those from earthquakes and other natural disasters (with tail indices that can be considerably less than one, see \citeauthor{ibragimov2015heavy}, \citeyear{ibragimov2015heavy}, and references therein).} 

The recent study by \cite{Taleb} provides (Hill's, see below) tail index estimates supporting extreme heavy-tailedness with $\zeta$ smaller than 1 and infinite first moments in the number of deaths from 72 major epidemic and pandemic diseases from 429 BC until the present. \cite{Toda1} report (Hill's) estimates of the tail index close to 1 implying infinite variances and first moments in the distribution of COVID-19 infections across the US counties at the beginning of the pandemic.

Several approaches to the inference about the tail index $\zeta$ of heavy-tailed distributions are available in the literature (see, among others, the reviews in \citeauthor{embrechts1997modelling}, \citeyear{embrechts1997modelling},  
\citeauthor{gabaix2011rank}, \citeyear{gabaix2011rank}, Ch. 3 in \citeauthor{ibragimov2015heavy}, \citeyear{ibragimov2015heavy}, and references therein). The two most commonly used ones are Hill’s
estimates and the OLS approach using the log-log rank-size regression.

It was reported in a number of studies that inference on the tail index using widely applied Hill’s estimates suffers from several problems, including sensitivity to dependence and small sample sizes (see, among others, Ch. 6 in \citeauthor{embrechts1997modelling}, \citeyear{embrechts1997modelling}). Motivated by these problems, several studies have focused on alternative approaches to the tail index estimation. For instance, \cite{huisman2001tail}  propose a weighted analogue of Hill’s estimator that is reported to correct its small sample bias for sample sizes less than 1,000. Using extreme value theory, \cite{MullerWang} focus on inference on the quantiles and tail probabilities of heavy-tailed variables with a fixed number $k$ of their extreme observations (order statistics) employed in estimation as is typical in relatively small samples of fat-tailed data. \cite{embrechts1997modelling}, among others, advocate sophisticated nonlinear procedures for tail index estimation.

\cite{gabaix2011rank} focus on econometrically justified inference on the tail index $\zeta$ in heavy-tailed power law models (\ref{power}) using the popular and widely applied approach based on log-log rank-size regressions $\log(Rank)=a-b \log(Size),$ with $b$ taken as an estimate of $\zeta.$ The reason for popularity of the approach is its simplicity and robustness. \cite{gabaix2011rank} provide a simple remedy for the inherent small sample bias in log-log rank-size approaches to inference on tail indices, and propose using the (optimal) shifts of 1/2 in ranks, with the tail index estimated by the parameter $b$ in (small sample bias-corrected) regressions $\log(Rank-1/2)=a-b \log(Size).$ \cite{gabaix2011rank} further derive the correct standard errors on the tail exponent $\zeta$ in the log-log rank-size regression approaches. The standard error on $\zeta$ in the above log-log rank-size regressions is not the OLS standard error but is asymptotically $(2/k)^{1/2}\zeta,$ where $k$ is the number of extreme (the largest) observations on the heavy-tailed variable $X$ used in tail index estimation (see also Ch. 3 in \citeauthor{ibragimov2015heavy}, \citeyear{ibragimov2015heavy}). The numerical results in
\cite{gabaix2011rank} point to advantages of the proposed approaches to inference on tail indices, including their robustness to dependence in the data and deviations from exact power laws in the form of slowly varying functions.

Naturally, it is important that inference on the tail index in heavy-tailed power law distributions (\ref{power}) is based on i.i.d. or stationary observations (op. cit.). The results in the previous section point to potential non-stationarity of the time series $\Delta I_t, \Delta D_t$ of daily COVID-19 infections and deaths and stationarity of their differences $\Delta^2 I_t, \Delta^2 D_t$ (the daily changes in the number of daily COVID-19 infections and deaths) in most of the countries considered. We, therefore, focus on estimation of the tail indices $\zeta$ in heavy-tailed power law models for the differences $\Delta^2 I_t$ and $\Delta^2 D_t.$ The implied tail indices for daily COVID-19 infections and deaths $\Delta I_t, \Delta D_t$ equal to the same values as in the case of $\Delta^2 I_t$ and $\Delta^2 D_t,$ as the former time series are cumulations of the latter ones.

Figures \ref{Fig1} and \ref{Fig2} provide the plots of Hill's estimates of the tail indices $\zeta$ in power law distributions for the time series $\Delta^2 I_t$ and $\Delta^2 D_t$ of daily changes in the number of COVID-19 infections and deaths in several countries considered (Australia, China, France, India, Italy, Russia, the UK and the US) with different number $k$ of extreme (largest) observations used in tail index estimation (the so-called Hill's plots, see Ch. 6 in \citeauthor{embrechts1997modelling}, \citeyear{embrechts1997modelling}, and also \citeauthor{Taleb}, \citeyear{Taleb}, for similar plots employed in the analysis of the inverse $\theta=1/\zeta$ of the tail index $\zeta$ in power law models (\ref{power}) for the number of deaths from major epidemic and pandemic diseases from ancient times until the present).\footnote{In the diagrams, $k$ plotted on the OX axis denotes the tail truncation level - the number of extreme observations - used in tail index estimation, in \% of the total sample size $N,$ that is, $k$ equals 2.5-15\% of the total sample size $N.$} Similarly, Figures \ref{Fig3} and \ref{Fig4} provide the log-log rank-size regression estimates of the tail indices $\zeta$ with optimal shifts $1/2$ in ranks proposed in \cite{gabaix2011rank} for the time series $\Delta^2 I_t$ and $\Delta^2 D_t$ in the above countries that use different truncation levels $k$ for the largest observations used in inference (see \citeauthor{IIK}, \citeyear{IIK}, for the analysis of such log-log rank-size plots for foreign exchange rates in emerging economies). The plots in Figures \ref{Fig1}-\ref{Fig4} also provide the corresponding 95\% confidence intervals for tail indices $\zeta$ in power law models (\ref{power}) for the changes in the number of daily COVID-19 infections and deaths.

The analysis of Figures \ref{Fig1}-\ref{Fig4} 
indicates that Hill's and log-log rank-size regression tail index estimates for the time series $\Delta^2 I_t$ and $\Delta^2 D_t$ of daily changes in the number of COVID-19 infections and deaths tend to stabilize as a sufficient number $k$ of extreme (largest) observations (order statistics) on $\Delta^2 I_t$ and $\Delta^2 D_t$ is used in inference. As expected, log-log rank-size regression estimates tend to be less sensitive to the choice of $k$ compared to Hill's estimates.

Importantly, the left-end points of the confidence intervals for tail indices $\zeta$ in power law models for daily changes in COVID-19 infections and deaths calculated using different tail truncation levels $k$ in the countries in the diagrams tend to be less than two indicating possibly infinite second moments and variances. Further, from the analysis of Figures \ref{Fig1}-\ref{Fig4} it follows that the tail indices may be even less than one for some of the countries indicating extreme heavy-tailedness with possibly infinite first moments.

The conclusions from heavy-tailedness analysis for other countries considered in the paper are similar to those above.


The conclusions on heavy-tailedness in the time series of daily COVID-19 infections and deaths are important as, according to the above discussion, they point to high likelihood of observing their large values. They further emphasise the necessity in the use of robust methods in statistical analysis and forecasting of the dynamics of the COVID-19 pandemic and its impacts, including the approaches robust to the problems of heavy-tailedness and heterogeneity in the data.

\subsection{Predictive regressions} \label{pred1}

This section presents the main results of the paper on statistically justified and robust evaluation of the effects of the COVID-19 pandemic on financial markets in different countries across the World. We focus on the analysis of predictive regressions for returns on major stock indices in the countries considered (see the introduction) 
incorporating the time series characterizing the infection and death rates from COVID-19 in the countries. Importantly, due to the problems of nonstationarity and the unit root dynamics in the time series $\Delta I_t, \Delta D_t$ of daily COVID-19 infections and COVID-19 related deaths in most of the countries discussed in Section \ref{UR}, estimation of the predictive regressions is provided for regression models for stock index returns $R_t$ with both the lagged daily infections/deaths $\Delta I_{t-1}, \Delta D_{t-1}$ and their (stationary) lagged differences $\Delta^2 I_{t-1}, \Delta^2 D_{t-1}$ (the daily changes in the number of COVID-19 infections and deaths) used as regressors. 

More precisely, the estimation results are provided for predictive regressions in the form
\begin{eqnarray} \label{pred}
R_{t} = \alpha+\beta X_{t-1}+\varepsilon_{t}, \end{eqnarray}
where $R_{t}$ are the excess returns on major stock indices in the countries considered at the end of the day $t$ given by the difference between the end of the day-$t$ stock index returns and the countries' interest rates (see Section \ref{data}), and the regressors $X_{t-1}$ are either the number $\Delta I_{t-1}, \Delta D_{t-1}$ of COVID-19 infections/deaths on day $t-1$ in the countries dealt with or the daily changes $\Delta^2 I_{t-1}=\Delta I_{t-1}-\Delta I_{t-2},$ $\Delta^2 D_{t-1}=\Delta D_{t-1}-\Delta D_{t-2}$ in the number of infections/deaths.\footnote{The well-known stylized fact of absence of linear autocorrelations in daily financial returns (see \cite{cont2001empirical} and references therein) implies exogeneity of regressors in the predictive regressions considered.}\footnote{We also conduct the analysis similar to that in the paper for the first and second differences of logarithms of daily infections and deaths. It points to unit root non-stationarity in the first log differences and the implied stationarity in the second log differences, similar to the differences of daily infections/deaths in Section \ref{UR}. We further obtain estimates in analogues of predictive regressions (\ref{pred}) with the lagged first and the (stationary) second differences of logarithms of daily infections and deaths used as regressors. The conclusions from the estimates of such regressions are mostly similar to those for regressions with the above differences $\Delta I_t, \Delta D_t$ $\Delta^2 I_t, \Delta^2 D_t$ of daily infections/deaths used as regressors. The estimation results are available on request.}


In order to account for autocorrelation and heteroskedasticity in the regressors and the error terms in predictive regressions (\ref{pred}) we use the widely applied HAC based methods (the standard errors and $t-$statistics with the quadratic spectral - QS - kernel and automatic choice of bandwidth as in \citeauthor{andrews1991heteroskedasticity}, \citeyear{andrews1991heteroskedasticity}, and the corresponding $p-$values based on standard normal approximations) in the analysis of statistical significance of the regression coefficients.

It is well known, however, that commonly used HAC inference methods and related approaches based on consistent standard errors often have poor finite sample properties, especially in the case of pronounced dependence, heterogeneity and heavy-tailedness in the data (see the discussion and the analysis in \citeauthor{IM}, \citeyear{IM, IM1}, Section 3.3 in \citeauthor{ibragimov2015heavy}, \citeyear{ibragimov2015heavy}, and references therein). To account for these problems, we also provide the analysis of statistical significance of predictive regression coefficients using the $t-$statistic approaches to robust inference under dependence, heterogeneity and heavy-tailedness of largely unknown form recently developed in \citeauthor{IM} (\citeyear{IM, IM1}). Following the approaches, robust large sample inference on a parameter of interest (e.g., a predictive regression coefficient $\beta$) is conducted as follows: The data is partitioned into a fixed number $q\ge 2$ (e.g., $q=2, 4, 8$) of groups, the model is estimated for each group, and inference is based on a standard $t-$test with the resulting $q$ parameter estimates.

In the context of inference on the coefficient $\beta$ in time series predictive regressions (\ref{pred}), the regression is estimated for $q$ groups of consecutive time series observations with $(j-1)T/q<t\le jT/q,$ $j=1, ..., q,$ resulting in $q$ group estimates $\hat{\beta}_j,$ $j=1, ..., q.$ The robust test of a hypothesis on the parameter $\beta$ is based on the $t-$statistic in the group OLS regression estimates $\hat{\beta}_j,$ $j=1, ..., q.$ E.g., the robust test of the null hypothesis $H_0: \beta=0$ against alternative $H_a: \beta\neq 0$ is based on the $t-$statistic $t_{\beta}=\sqrt{q}\frac{\overline{\hat{\beta}}}{s_{\hat{\beta}}},$ where $\overline{\hat{\beta}}=q^{-1}\sum_{j=1}^q \hat{\beta}_j$ and $s_{\hat{\beta}}^2=(q-1)^{-1}\sum_{j=1}^ q (\hat{\beta}_j-\overline{\hat{\beta}})^2.$ The above null hypothesis $H_0$ is rejected in favor of the alternative $H_a$ at level $\alpha\le 8.3\%$ (e.g., at the usual significance level $\alpha=5\%$) if the absolute value $|t_{\hat{\beta}}|$ of the $t-$statistic in group estimates $\hat{\beta}_j$ exceeds the $(1-\alpha/2)-$quantile of the standard Student-$t$ distribution with $q-1$ degrees of freedom.\footnote{One-sided tests are conducted in a similar way, and the approaches further provide robust confidence intervals for the unknown parameters $\beta$ (\citeauthor{IM1}, \citeyear{IM1}, \citeauthor{ibragimov2015heavy}, \citeyear{ibragimov2015heavy}).}

The $t-$statistic based approaches do not require at all estimation of limiting variances of estimators of interest. As discussed in \cite{IM1, ibragimov2015heavy, IM}, they result in asymptotically valid inference under the assumptions that the group estimators of a parameter of interest are asymptotically independent, unbiased and Gaussian of possibly different variances.\footnote{Justification of asymptotic validity of the robust $t-$statistic inference approaches in \cite{IM, IM1} is based on a small sample result in \cite{bakirov2006student} that implies validity of the standard $t-$test under independent heterogeneous Gaussian observations and its analogues for two-sample $t-$tests obtained in \cite{IM1}.} The assumptions are satisfied in a wide range of econometric models and dependence, heterogeneity and heavy-tailedness settings of a largely unknown type. The numerical analysis in \cite{IM, IM1, ibragimov2015heavy} indicates favorable finite sample performance of the $t-$statistic based robust inference approaches in inference on models with time series, panel, clustered and spatially correlated data.\footnote{See also \cite{esarey} for a detailed numerical analysis of finite sample performance of different inference procedures, including $t-$statistic approaches,  under small number of clusters of dependent data  and their software (STATA and R) implementation.}\footnote{The \textit{t-}statistic robust inference approach proposed in \cite{IM} provides a formal justification for the widespread Fama–MacBeth method for inference in
panel regressions with heteroskedasticity (see \citeauthor{FM}, \citeyear{FM}). Following the method, one estimates the regression separately for each year, and then tests
hypotheses about the coefficient of interest using the \textit{t-}statistic of the resulting yearly coefficient estimates. The Fama–MacBeth approach is a special case of the \textit{t-}statistic based approach to inference, with observations of the same year collected in a group.}\footnote{See, among others, \cite{Bloom}, \cite{Krueger}, \cite{BW}, \cite{verner}, \cite{ChenRI} and \cite{Tim} for empirical applications of the robust $t-$statistic  inference approaches in \cite{IM, IM1}.} Importantly, the $t-$statistic based approaches to robust inference may also be used under convergence of group estimators of a parameter interest to scale mixtures of normal distributions as in the case of models under heavy-tailedness with infinite variances and in regressions with non-stationary exogenous regressors. \footnote{See Section 3.3.3 in \cite{ibragimov2015heavy} for applications of the robust $t-$statistic approaches in inference in infinite variance heavy-tailed models. The recent works by \cite{Anat}, \cite{Pedersen} and \cite{IKP} provide further applications of the approaches in robust inference on general classes of GARCH and AR-GARCH-type models exhibiting heavy-tailedness and volatility clustering properties typical for real-world financial returns, foreign exchange rates and other important economic and financial time series. The recent paper by \cite{IKS} focuses on applications of the $t-$statistic approaches in inference on predictive regressions with persistent and/or fat-tailed regressors and errors. }

Tables \ref{tab2} and \ref{tab3} provide the results of the assessment of statistical significance of the coefficients $\beta$ on the lagged time series $\Delta I_{t-1}, \Delta D_{t-1}$ of daily COVID-19 infections/deaths and their differences $\Delta^2 I_{t-1}, \Delta^2 D_{t-1}$  - the daily changes in the number of infections and deaths from the disease - in predictive regressions (\ref{pred}) for the countries considered. More precisely, the tables provide the values of HAC $t-$statistic with the QS kernel and the automatic choice of bandwidth discussed above as well as the values of the $t-$statistic $t_{\beta}$ in estimates $\hat{\beta}_j,$ $j=1, ..., q,$ of the slope parameter $\beta$ obtained using $q=4, 8, 12$ and 16 groups of consecutive time series observations. The asterisks in the tables indicate statistical significance of the slope coefficient (*** for significance at 1\%, ** for significance at 5\% and * for significance at 10\%) implied by formal comparisons of the HAC $t-$statistics with the quantiles of a standard normal distribution. As described above, following the $t-$statistic approaches to robust inference in \cite{IM, IM1}, (the absence of) statistical significance of the slope coefficient $\beta$ is assessed using the comparisons of the $t-$statistic in group estimates of the regression coefficient in the table with the  quantiles of Student-$t$ distributions with $q-1$ degrees of freedom.

The values of HAC $t-$statistics in Tables \ref{tab2} and \ref{tab3} indicate an apparently spurious statistical significance of the (potentially non-stationary) lagged daily infections and deaths $\Delta I_{t-1}, \Delta D_{t-1}$ from COVID-19 in predictive regressions for returns on the main stock indices in some countries. In the case of daily COVID-19 infections, this is observed for the UK, Italy, Russia, the US, Brazil, China, South Korea and Australia, and, in the case of daily COVID-19 related deaths, for India, Canada, Brazil, Mexico, China, South Korea, Indonesia and Australia. The coefficients $\Delta I_{t-1}, \Delta D_{t-1}$ whose significance is indicated by the HAC $t-$statistics have the expected negative sign.

According to the HAC $t-$statistics in the tables, in the case of (econometrically justified) predictive regressions for excess returns on major stock indices considered involving (stationary) differences $\Delta I^2_{t-1},$ $\Delta D^2_{t-1}$  of the lagged daily COVID-19 infections and deaths (daily changes in the number of infections and deaths from the disease), the coefficients at the regressors $\Delta I^2_{t-1},$ $\Delta D^2_{t-1}$ are not statistically significant at conventional levels for most of the countries. Exceptions are, in the case of daily changes $\Delta I^2_{t-1}$ in the number of COVID-19 infections, are Italy, India, Brazil and Argentina, where the coefficient at the regressor $\Delta I^2_{t-1}$ is highly significant (at 1\% level) according to the HAC $t-$statistics, with the expected negative sign, and also the UK where some significance (at 10\% level) is observed, albeit with the positive sign at the coefficient. In the case of predictive regressions incorporating the daily changes $\Delta D^2_{t-1}$ in the number of COVID-19 related deaths, the HAC $t-$statistics indicate high statistical significance (at 1\%) of the coefficient at the regressor $\Delta D^2_{t-1}$ only for the UK, India, Brazil, Mexico, South Korea and Australia; among these, the predictive regression coefficient has the expected negative sign only in the case of Australia. 


However, the lagged daily COVID-19 infection and death rates $\Delta I_{t-1}$ and $\Delta D_{t-1}$ and their (stationary) differences $\Delta^2 I_{t-1}$ and $\Delta^2 D_{t-1}$  appear not to be statistically significant in predictive regressions for stock index returns in all countries considered according to the (econometrically justified) robust $t-$statistic approaches with different choices of the number of groups $q.$ 

\section{Conclusion} \label{conclude}
This paper presented the results of theoretically justified and robust statistical analysis of the effects of the COVID-19 pandemic on financial markets in different countries across the World. The analysis is based on robust inference in predictive regressions for the returns on the countries' major stock indices incorporating the time series characterizing the dynamics in the COVID-19 related deaths rates.

The paper further presented the results of the statistical analysis of (non-)stationarity, heavy-tailedness and tail risk in the time series on infections/death rates from COVID-19 in the countries considered. The obtained results point to non-stationary unit root dynamics and pronounced heavy-tailedness with possibly infinite variances and fist moments in the time series of daily COVID-19 infections and deaths in most of the countries dealt with.

According to the results in the paper, the standard HAC inference methods indicate apparently spurious statistical significance of the (potentially non-stationary) lagged daily infections/deaths from COVID-19 in predictive regressions for returns on the major stock indices in some countries. HAC methods also point to  statistical significance of (stationary) lagged daily changes in the number of COVID-19 infections and deaths for some countries. For daily changes in COVID-19 infections, high statistical significance, with the expected negative signs of the predictive regression coefficients, is observed In the case of Italy, India, Brazil and Argentina.

Motivated by the results on high persistence and heavy-tailedness in daily COVID-19 infections/ deaths time series obtained in the paper and also by poor finite sample properties of HAC inference methods, we further provide the analysis of statistical significance of the coefficients in the predictive regressions using the recently developed $t-$statistic approaches to robust inference. The analysis using the robust $t-$statistic inference approaches indicates that the lagged daily COVID-19 death rates and their (stationary) differences appear to be statistically insignificant in predictive regressions for stock index returns in essentially all countries considered in the analysis.

The analysis and conclusions in the paper emphasize the necessity in the use of robust inference methods accounting for autocorrelation, heterogeneity and heavy-tailedness in statistical and econometric analysis and forecasting of key time series and variables related to the COVID-19 pandemic and its effects on economic and financial markets and society. They further emphasize the importance of the use of correctly specified models of the COVID-19 pandemic and its effects incorporating stationary time series and variables such as the daily changes in COVID-19 related deaths used in predictive regressions in this work.

Further research may focus on robust 
tests for structural breaks in models of the dynamics of the pandemic and its effects on financial and economic markets using the robust inference approaches based on two-sample $t-$statistics in \cite{IM1}, and applications of inference methods such as sign- and rank-based tests that are robust to relatively small sample sizes of observations in statistical analysis of key models related to the spread of COVID-19. It would also be of interest to apply further estimation approaches for heavy-tailed models for time series associated with the pandemic that are robust to small samples, including 
the recently developed fixed-$k$ inference approaches for power-law models (\ref{power}) in \cite{MullerWang}. The analysis in these directions is currently under way by the authors and co-authors.


\bibliographystyle{agsm} 
\bibliography{reportbb}

\restoregeometry
\makeatletter
\let\origsection\section
\renewcommand\section{\@ifstar{\starsection}{\nostarsection}}

\newcommand\nostarsection[1]
{\sectionprelude\origsection{#1}\sectionpostlude}

\newcommand\starsection[1]
{\sectionprelude\origsection*{#1}\sectionpostlude}

\newcommand\sectionprelude{%
  \vspace{-1em}
}

\newcommand\sectionpostlude{%
  \vspace{-1em}
}
\makeatother


\appendix

\section{Tables}\label{sec:tables}

\begin{table}[H]
\centering
\scriptsize
\caption{Wild bootstrap quasi-differenced unit root tests for Infections based on Rademacher distribution with sieve based recolouring ($p-$values in brackets) \label{tab1}}
\begin{tabularx}{1\textwidth}{ccccccc|cccccc} \toprule
&\multicolumn{6}{c}{$\Delta Infections$}&\multicolumn{6}{c}{$\Delta^2 Infections$}\\\hline
&$LR$&$MZ_\alpha$&$MSB$&$MZ_t$&$MP_t$&$ADF$&$LR$&$MZ_\alpha$&$MSB$&$MZ_t$&$MP_t$&$ADF$\\\hline
UK&0.44&-2.04&0.49&-1.00&11.96&-0.99&120.41&-202.64&0.05&-10.07&0.12&-23.98\\
&(0.42)&(0.48)&(0.67)&(0.42)&(0.52)&(0.43)&(0.00)&(0.00)&(0.00)&(0.00)&(0.00)&(0.00)\\
Germany&0.59&-2.98&0.39&-1.17&8.11&-1.13&136.23&-190.84&0.05&-9.75&0.16&-27.93\\
&(0.44)&(0.44)&(0.54)&(0.39)&(0.45)&(0.44)&(0.00)&(0.00)&(0.00)&(0.00)&(0.00)&(0.00)\\
France&0.34&7.63&0.34&2.62&25.40&-0.95&147.60&-171.55&0.05&-8.77&0.85&-31.02\\
&(0.41)&(1)&(0.23)&(1)&(0.73)&(0.38)&(0.00)&(0.00)&(0.00)&(0.00)&(0.00)&(0.00)\\
Italy&0.00&-0.26&0.78&-0.20&34.90&0.30&89.82&-204.64&0.05&-10.01&0.26&-17.77\\
&(0.64)&(0.6)&(0.91)&(0.59)&(0.82)&(0.73)&(0.00)&(0.00)&(0.00)&(0.00)&(0.00)&(0.00)\\
Spain&6.41&-12.76&0.19&-2.48&2.12&-3.12&&&&&&\\
&(0.01)&(0.05)&(0.05)&(0.06)&(0.05)&(0.02)&&&&&&\\
Russia&0.01&-0.22&1.09&-0.24&61.72&-0.24&140.35&-182.19&0.05&-9.54&0.13&-29.47\\
&(0.6)&(0.65)&(0.93)&(0.6)&(0.88)&(0.6)&(0.00)&(0.00)&(0.00)&(0.00)&(0.00)&(0.00)\\
Netherland&0.00&-0.07&0.70&-0.05&30.29&0.03&94.31&-194.36&0.05&-9.83&0.16&-19.16\\
&(0.7)&(0.71)&(0.88)&(0.71)&(0.83)&(0.73)&(0.00)&(0.00)&(0.00)&(0.00)&(0.00)&(0.00)\\
Sweden&12.97&-22.88&0.15&-3.38&1.07&-4.37&&&&&&\\
&(0)&(0.03)&(0.04)&(0.03)&(0.02)&(0.01)&&&&&&\\
India&0.00&-0.04&0.95&-0.04&50.30&-0.01&132.27&-193.08&0.05&-9.81&0.15&-26.94\\
&(0.64)&(0.67)&(0.93)&(0.66)&(0.88)&(0.67)&(0.00)&(0)&(0.10)&(0)&(0)&(0.00)\\
Austria&0.01&-0.71&0.62&-0.44&22.17&-0.25&90.76&-194.57&0.05&-9.86&0.13&-18.44\\
&(0.63)&(0.59)&(0.75)&(0.56)&(0.68)&(0.63)&(0.00)&(0.00)&(0.00)&(0.00)&(0.00)&(0.00)\\
Finland&0.00&-0.24&0.56&-0.13&21.32&0.89&137.87&-187.21&0.05&-9.66&0.15&-28.53\\
&(0.72)&(0.71)&(0.8)&(0.7)&(0.75)&(0.88)&(0.00)&(0.00)&(0.00)&(0.00)&(0.00)&(0.00)\\
Ireland&4.71&-25.97&0.14&-3.60&0.96&-3.36&&&&&&\\
&(0.02)&(0.03)&(0.03)&(0.03)&(0.03)&(0.03)&&&&&&\\
US&0.14&-0.80&0.74&-0.59&27.91&-0.59&146.30&-181.65&0.05&-9.52&0.15&-30.76\\
&(0.55)&(0.63)&(0.92)&(0.55)&(0.81)&(0.55)&(0.00)&(0.00)&(0.00)&(0.00)&(0.00)&(0.00)\\
Lithuania&0.30&-1.42&0.59&-0.83&17.01&-0.82&112.89&-187.92&0.05&-9.69&0.13&-23.35\\
&(0.49)&(0.54)&(0.77)&(0.47)&(0.65)&(0.49)&(0.00)&(0.00)&(0.00)&(0.00)&(0.00)&(0.00)\\
Canada&0.30&-1.42&0.59&-0.83&17.01&-0.82&135.25&-192.72&0.05&-9.79&0.17&-27.59\\
&(0.49)&(0.54)&(0.77)&(0.47)&(0.65)&(0.49)&(0.00)&(0.00)&(0.00)&(0.00)&(0.00)&(0.00)\\
Brazil&0.04&-0.18&0.62&-0.11&25.12&-0.37&128.47&-174.91&0.05&-9.35&0.14&-27.54\\
&(0.6)&(0.66)&(0.84)&(0.65)&(0.76)&(0.6)&(0.00)&(0.00)&(0.00)&(0.00)&(0.00)&(0.00)\\
Mexico&0.00&-1.20&0.49&-0.59&14.86&-0.13&128.34&-173.32&0.05&-9.31&0.14&-27.64\\
&(0.68)&(0.5)&(0.6)&(0.5)&(0.53)&(0.68)&(0.00)&(0.00)&(0.00)&(0.00)&(0.00)&(0.00)\\
Argentina&0.62&-2.83&0.42&-1.19&8.65&-1.18&96.23&-191.99&0.05&-9.77&0.16&-19.69\\
&(0.32)&(0.33)&(0.47)&(0.29)&(0.34)&(0.31)&(0.00)&(0.00)&(0.00)&(0.00)&(0.00)&(0.00)\\
Japan&0.01&-0.34&0.86&-0.29&39.38&-0.23&90.89&-208.23&0.05&-10.19&0.13&-17.85\\
&(0.7)&(0.69)&(0.94)&(0.65)&(0.89)&(0.69)&(0.00)&(0.00)&(0.00)&(0.00)&(0.00)&(0.00)\\
China&0.34&-1.82&0.51&-0.94&13.17&-0.89&&&&&&\\
 &(0.00)&(0.00)&(0.00)&(0.00)&(0.00)&(0.00)&&&&&&\\
South Korea&6.73&-23.11&0.15&-3.40&1.06&-3.95&131.66&-199.96&0.05&-9.99&0.14&-26.39\\
&(0.08)&(0.04)&(0.04)&(0.04)&(0.04)&(0.04)&(0.00)&(0.00)&(0.00)&(0.00)&(0.00)&(0.00)\\
Indonesia&0.59&-3.36&0.36&-1.22&7.26&-1.17&113.81&-185.58&0.05&-9.59&0.19&-23.67\\
&(0.24)&(0.23)&(0.28)&(0.23)&(0.23)&(0.24)&(0.00)&(0.00)&(0.00)&(0.00)&(0.00)&(0.00)\\
Australia&0.00&0.17&1.08&0.19&66.41&0.22&154.91&-163.02&0.06&-9.03&0.15&-34.39\\
&(0.71)&(0.75)&(0.99)&(0.78)&(0.97)&(0.78)&(0.00)&(0.00)&(0.00)&(0.00)&(0.00)&(0.00)\\
  \bottomrule
\end{tabularx}
\end{table}

\pagebreak

\begin{table}[H]
\centering
\scriptsize
\caption{Wild bootstrap quasi-differenced unit root tests for Deaths based on Rademacher distribution with sieve based recolouring ($p-$values in brackets) \label{tab11}}
\begin{tabularx}{1\textwidth}{ccccccc|cccccc} \toprule
&\multicolumn{6}{c}{$\Delta Deaths$}&\multicolumn{6}{c}{$\Delta^2 Deaths$}\\\hline
&$LR$&$MZ_\alpha$&$MSB$&$MZ_t$&$MP_t$&$ADF$&$LR$&$MZ_\alpha$&$MSB$&$MZ_t$&$MP_t$&$ADF$\\\hline
UK&0.38&-1.58&0.56&-0.89&15.55&-0.92&100.54&-189.85&0.05&-9.74&0.13&-20.69\\
&(0.35)&(0.39)&(0.53)&(0.34)&(0.45)&(0.35)&(0.00)&(0.00)&(0.00)&(0.00)&(0.00)&(0.00)\\
Germany&0.65&-2.44&0.43&-1.06&9.77&-1.19&116.42&-178.45&0.05&-9.43&0.16&-24.70\\
&(0.39)&(0.42)&(0.59)&(0.38)&(0.46)&(0.38)&(0.00)&(0.00)&(0.00)&(0.00)&(0.00)&(0.00)\\
France&1.70&-4.15&0.29&-1.22&6.22&-1.77&141.15&-165.54&0.05&-9.01&0.27&-30.95\\
&(0.1)&(0.17)&(0.13)&(0.22)&(0.18)&(0.12)&(0.00)&(0.00)&(0.00)&(0.00)&(0.00)&(0.00)\\
Italy&0.10&-0.92&0.59&-0.54&19.49&-0.54&119.48&-188.42&0.05&-9.69&0.15&-24.67\\
&(0.5)&(0.49)&(0.61)&(0.48)&(0.54)&(0.5)&(0.00)&(0.00)&(0.00)&(0.00)&(0.00)&(0.00)\\
Spain&4.49&-8.30&0.22&-1.84&3.69&-2.53&&&&&&\\
&(0.01)&(0.04)&(0.03)&(0.07)&(0.07)&(0.01)&&&&&&\\
Russia&0.00&0.28&1.19&0.33&82.22&0.52&110.55&-176.31&0.05&-9.39&0.14&-23.60\\
&(0.67)&(0.76)&(0.97)&(0.8)&(0.95)&(0.86)&(0.00)&(0.00)&(0.00)&(0.00)&(0.00)&(0.00)\\
Netherland&0.68&-2.87&0.42&-1.20&8.53&-1.22&120.78&-176.98&0.05&-9.41&0.14&-25.65\\
&(0.24)&(0.27)&(0.34)&(0.23)&(0.28)&(0.24)&(0.00)&(0.00)&(0.00)&(0.00)&(0.00)&(0.00)\\
Sweden&8.45&-10.10&0.22&-2.25&2.43&-3.21&&&&&&\\
&(0.01)&(0.03)&(0.03)&(0.03)&(0.02)&(0.01)&&&&&&\\
India&0.14&-0.74&0.80&-0.59&31.76&-0.60&142.49&-137.18&0.06&-8.28&0.18&-34.50\\
&(0.43)&(0.49)&(0.69)&(0.44)&(0.62)&(0.45)&(0.00)&(0)&(0.10)&(0)&(0)&(0.00)\\
Austria&0.57&-2.61&0.43&-1.13&9.35&-1.12&125.60&-165.39&0.05&-9.09&0.15&-27.69\\
&(0.4)&(0.43)&(0.57)&(0.38)&(0.46)&(0.4)&(0.00)&(0.00)&(0.00)&(0.00)&(0.00)&(0.00)\\
Finland&7.47&-14.26&0.19&-2.65&1.78&-3.33&&&&&&\\
&(0)&(0.03)&(0.03)&(0.02)&(0.02)&(0.01)&&&&&&\\
Ireland&2.72&-7.31&0.26&-1.91&3.35&-2.17&128.63&-161.97&0.06&-9.00&0.15&-28.66\\
&(0.07)&(0.06)&(0.08)&(0.05)&(0.05)&(0.05)&(0.00)&(0.00)&(0.00)&(0.00)&(0.00)&(0.00)\\
US&0.13&-0.86&0.75&-0.64&27.58&-0.61&81.99&-189.13&0.05&-9.72&0.13&-16.91\\
&(0.49)&(0.51)&(0.84)&(0.45)&(0.74)&(0.5)&(0.00)&(0.00)&(0.00)&(0.00)&(0.00)&(0.00)\\
Lithuania&0.18&-1.09&0.64&-0.70&20.83&-0.68&134.84&-137.92&0.06&-8.30&0.18&-32.56\\
&(0.54)&(0.6)&(0.85)&(0.55)&(0.72)&(0.55)&(0.00)&(0.00)&(0.00)&(0.00)&(0.00)&(0.00)\\
Canada&0.28&-1.36&0.59&-0.80&17.37&-0.83&135.84&-152.09&0.06&-8.71&0.17&-31.22\\
&(0.37)&(0.44)&(0.6)&(0.38)&(0.52)&(0.37)&(0.00)&(0.00)&(0.00)&(0.00)&(0.00)&(0.00)\\
Brazil&0.00&-0.75&0.52&-0.39&17.28&0.44&120.57&-167.56&0.05&-9.15&0.15&-26.41\\
&(0.69)&(0.51)&(0.57)&(0.53)&(0.51)&(0.84)&(0.00)&(0.00)&(0.00)&(0.00)&(0.00)&(0.00)\\
Mexico&1.18&-3.75&0.36&-1.37&6.53&-1.40&125.87&-158.55&0.06&-8.90&0.15&-28.35\\
&(0.18)&(0.22)&(0.32)&(0.19)&(0.22)&(0.22)&(0.00)&(0.00)&(0.00)&(0.00)&(0.00)&(0.00)\\
Argentina&4.83&-9.49&0.23&-2.17&2.63&-2.56&&&&&&\\
&(0.03)&(0.05)&(0.05)&(0.05)&(0.05)&(0.05)&&&&&&\\
Japan&0.52&-2.21&0.43&-0.96&10.42&-1.07&147.75&-156.80&0.06&-8.84&0.17&-33.43\\
&(0.4)&(0.44)&(0.57)&(0.4)&(0.46)&(0.39)&(0.00)&(0.00)&(0.00)&(0.00)&(0.00)&(0.00)\\
China&15.18&-44.19&0.11&-4.70&0.55&-5.97&&&&&&\\
 &(0.00)&(0.00)&(0.00)&(0.00)&(0.00)&(0.00)&&&&&&\\
South Korea&1.03&-3.33&0.37&-1.23&7.32&-1.39&152.66&-152.24&0.06&-8.68&0.23&-33.43\\
&(0.2)&(0.27)&(0.31)&(0.25)&(0.27)&(0.2)&(0.00)&(0.00)&(0.00)&(0.00)&(0.00)&(0.00)\\
Indonesia&0.09&-0.95&0.54&-0.52&17.57&-0.54&121.94&-171.88&0.05&-9.24&0.19&-26.12\\
&(0.56)&(0.6)&(0.78)&(0.57)&(0.7)&(0.55)&(0.00)&(0.00)&(0.00)&(0.00)&(0.00)&(0.00)\\
Australia&1.61&-5.04&0.31&-1.59&4.86&-1.78&135.87&-160.31&0.06&-8.95&0.15&-30.43\\
&(0.23)&(0.21)&(0.22)&(0.2)&(0.2)&(0.21)&(0.00)&(0.00)&(0.00)&(0.00)&(0.00)&(0.00)\\
  \bottomrule
\end{tabularx}
\end{table}

\pagebreak

\begin{table}[!ht]
\centering
\scriptsize
\caption{Predictive regression tests for infection rates\label{tab2}}
\begin{tabularx}{0.99\textwidth}{ccccccc|cccccc} \toprule
&\multicolumn{6}{c}{$\Delta Infections$}&\multicolumn{6}{c}{$\Delta^2 Infections$}\\\hline
&T&q=4&q=8&q=12&q=16&HAC&T&q=4&q=8&q=12&q=16&HAC\\\hline
UK FTSE 100&285&-0.18&1.59&0.45&0.45&-4.57$^{***}$&284&-0.67&-0.43&-0.31&-0.74&1.92$^{*}$\\
Germany DAX&288&-0.11&0.23&0.68&-0.03&-1.15&287&1.42&0.59&0.19&0.07&-0.15\\
France CAC 40&294&1.11&0.91&-0.11&0.20&-1.43&293&1.22&-0.10&-0.43&0.07&-0.79\\
Italy FTSE MIB&287&0.97&1.03&0.29&0.14&-2.78$^{***}$&286&-1.11&-1.45&-0.07&-1.16&-2.24$^{***}$\\
Spain IBEX 35&288&-1.35&1.39&1.02&2.45$^{**}$&-1.12&287&0.70&-0.63&-0.79&0.01&-0.14\\
Russia MOEX&280&-0.99&-1.03&-1.09&-0.92&-3.48$^{***}$&279&0.89&1.37&-0.62&1.28&-0.13\\
Netherland AEX&270&0.26&-0.23&-0.84&-1.05&-0.16&269&-0.60&-0.73&-0.95&-1.16&1.04\\
Sweden OMXS 30&281&0.43&-0.85&0.02&0.45&-1.51&280&-0.31&-0.85&-0.35&-1.77&-0.37\\
India SENSEX&280&0.86&1.44&1.43&-0.07&-4.25&279&0.34&-0.28&0.01&-0.01&-3.05$^{***}$\\
Austria ATX&268&1.41&0.57&0.40&-0.75&-0.89&267&0.39&-0.44&0.75&0.06&1.07\\
Finland OMX Helsinki 25&284&1.17&0.18&0.70&0.58&-1.48&283&-0.91&-1.41&0.47&-0.48&-0.22\\
Ireland ISEQ&267&0.17&1.20&-1.74&-0.21&-1.75&266&-1.23&-0.94&-1.66&-1.43&-0.50\\
US Dow Jones&291&1.43&1.81&-0.17&-0.11&-2.67$^{***}$&290&0.95&1.89&0.33&0.13&-1.90\\
US S\&P 500&291&1.63&2.25&0.62&1.35&-2.67$^{***}$&290&1.01&1.80&1.39&1.62&-1.94\\
Lithuania OMX Vilnius&264&1.23&0.24&0.70&0.83&-1.15&263&-0.51&-0.54&0.34&0.01&-1.54\\
Canada TSX&287&0.02&0.49&1.48&1.92&-3.53&286&3.07&1.26&1.53&1.04&0.73\\
Brazil iBovespa&259&-1.08&-1.21&-0.99&-0.85&-6.32$^{***}$&258&0.40&0.60&0.14&1.10&-2.14$^{**}$\\
Mexico IPC&263&-1.42&-1.56&-1.49&-1.05&-3.51&262&1.40&1.20&1.09&1.29&3.50\\
Argentina Merval&252&-0.85&0.92&-0.32&1.21&-1.73&251&0.64&-0.15&-0.90&-0.25&-3.14$^{***}$\\
Japan NIKKEI 225&281&1.27&3.27$^{***}$&1.12&1.18&-0.95&280&1.99&1.43&1.73&0.71&0.92\\
China SHANGHAI&277&0.77&-0.54&-0.22&1.34&-8.99$^{***}$&276&0.81&1.28&2.29$^{**}$& 2.00 &1.05\\
South KOSPI&284&0.91&1.02&1.06&1.40&-9.03$^{***}$&283&1.34&1.94&0.70&0.73&0.20\\
Indonesia JCI&251&-0.95&-0.58&-0.25&-0.56&-1.15&250&-0.67&-0.37&0.40&0.30&-0.40\\
Australia ASX 50&290&-1.19&-0.91&1.08&-0.89&-2.52$^{***}$&289&1.33&1.57&1.50&1.98&0.70\\
Australia ASX 200&290&-1.20&-0.91&1.12&-0.93&-2.52$^{***}$&289&1.21&1.56&1.39&1.91&0.81\\
Australian All&290&-1.20&-0.95&1.02&-1.01&-2.52$^{***}$&289&1.13&1.53&1.26&1.78&0.85\\
  \bottomrule
\end{tabularx}
\end{table}

\pagebreak

\begin{table}[!ht]
\centering
\scriptsize
\caption{Predictive regression tests for death rates\label{tab3}}
\begin{tabularx}{0.99\textwidth}{ccccccc|cccccc} \toprule
&\multicolumn{6}{c}{$\Delta Deaths$}&\multicolumn{6}{c}{$\Delta^2 Deaths$}\\\hline
&T&q=4&q=8&q=12&q=16&HAC&T&q=4&q=8&q=12&q=16&HAC\\\hline
UK FTSE 100&260&0.33&-0.45&1.24&-0.40&-1.72&259&1.53&1.30&0.49&1.48&5.13$^{***}$\\
Germany DAX&259&-0.87&-1.27&-1.24&-0.90&-0.55&258&-0.88&-0.84&-1.62&-1.05&-0.72\\
France CAC 40&278&2.37&0.53&1.43&-0.87&-0.83&277&0.81&0.37&0.48&-0.58&0.15\\
Italy FTSE MIB&272&-0.17&-0.54&0.71&0.33&-1.38&271&-0.90&-1.49&-0.74&-0.27&-0.24\\
Spain IBEX 35&267&1.05&0.37&1.31&1.24&-1.34&266&0.85&1.68&1.55&1.30&-0.99\\
Russia MOEX&248&-1.18&-1.15&-0.74&-0.24&-3.06&247&-1.17&0.21&0.16&0.88&-0.09\\
Netherland AEX&264&-0.73&-0.88&-1.68&-1.66&0.04&263&-0.74&-0.05&-0.90&-1.19&0.31\\
Sweden OMXS 30&255&-1.48&-1.68&-0.92&-1.60&1.29&254&-1.38&-1.48&0.19&-1.70&1.43\\
India SENSEX&253&0.24&1.25&1.34&1.48&-4.20$^{***}$&252&1.50&-0.24&-0.27&0.87&-1.70\\
Austria ATX&256&-0.55&-1.48&-1.13&-0.68&1.43&255&-0.74&-1.03&-1.28&-0.34&3.08$^{***}$\\
Finland OMX Helsinki 25&246&1.44&1.39&1.75&1.66&0.54&245&1.18&0.97&1.10&1.23&1.60\\
Ireland ISEQ&260&0.23&-0.32&0.39&-0.78&0.72&259&0.62&1.07&0.34&0.77&1.50\\
US Dow Jones&264&-0.47&-0.72&-0.17&0.41&-1.46&263&1.36&0.88&1.03&0.77&0.26\\
US S\&P 500&264&0.19&0.26&0.26&0.67&-1.46&263&1.40&1.04&1.01&0.87&0.25\\
Lithuania OMX Vilnius&248&3.77&-0.55&-0.83&-0.46&1.50&247&-1.19&-0.11&-1.35&-0.50&1.23\\
Canada TSX&258&0.92&1.29&0.79&0.29&-2.96$^{***}$&257&1.97&1.25&0.11&-0.49&0.21\\
Brazil iBovespa&246&-1.84&0.05&0.04&0.64&-3.31$^{***}$&245&1.40&1.32&0.89&1.43&4.26$^{***}$\\
Mexico IPC&250&-1.58&-1.89&-1.29&-1.75&-5.30$^{***}$&249&0.51&1.46&0.94&-0.83&4.00$^{***}$\\
Argentina Merval&248&-0.63&0.89&0.79&-0.81&-1.14&247&0.82&-0.44&1.04&1.36&-1.64\\
Japan NIKKEI 225&266&0.12&0.75&0.04&-0.25&-0.50&265&-0.94&-1.14&-1.67&-1.60&0.76\\
China SHANGHAI&277&0.01&1.52&-0.65&0.08&-6.25$^{***}$&276&0.68&1.75&0.80&1.21&0.37\\
South KOSPI&265&-1.40&-0.79&-0.60&1.08&-6.25$^{***}$&264&-1.11&0.11&0.48&0.29&12.03$^{***}$\\
Indonesia JCI&244&-0.99&-0.81&-0.95&-1.22&-3.39$^{***}$&243&-0.61&-0.47&-0.46&-0.94&-0.86\\
Australia ASX 50&266&-1.41&-1.06&-0.65&-0.94&-4.88$^{***}$&265&1.22&1.51&1.55&1.81&-4.99$^{***}$\\
Australia ASX 200&266&-1.39&-1.06&-0.52&-0.89&-4.82$^{***}$&265&1.25&1.59&1.65&1.79&-5.02$^{***}$\\
Australian All&266&-1.41&-1.07&-0.57&-0.91&-4.77$^{***}$&265&1.27&1.60&1.67&1.77&-5.18$^{***}$\\
  \bottomrule
\end{tabularx}
\end{table}

\pagebreak

\section{Figures} \label{figures}
\newgeometry{margin=2cm}
\begin{landscape}

\begin{figure}[!h]%
\caption{Hill's tail index estimates for positive changes in daily COVID-19 infections}\label{Fig1}
\begin{center}%
\subfigure[Australia]{\includegraphics[width=0.4\linewidth]{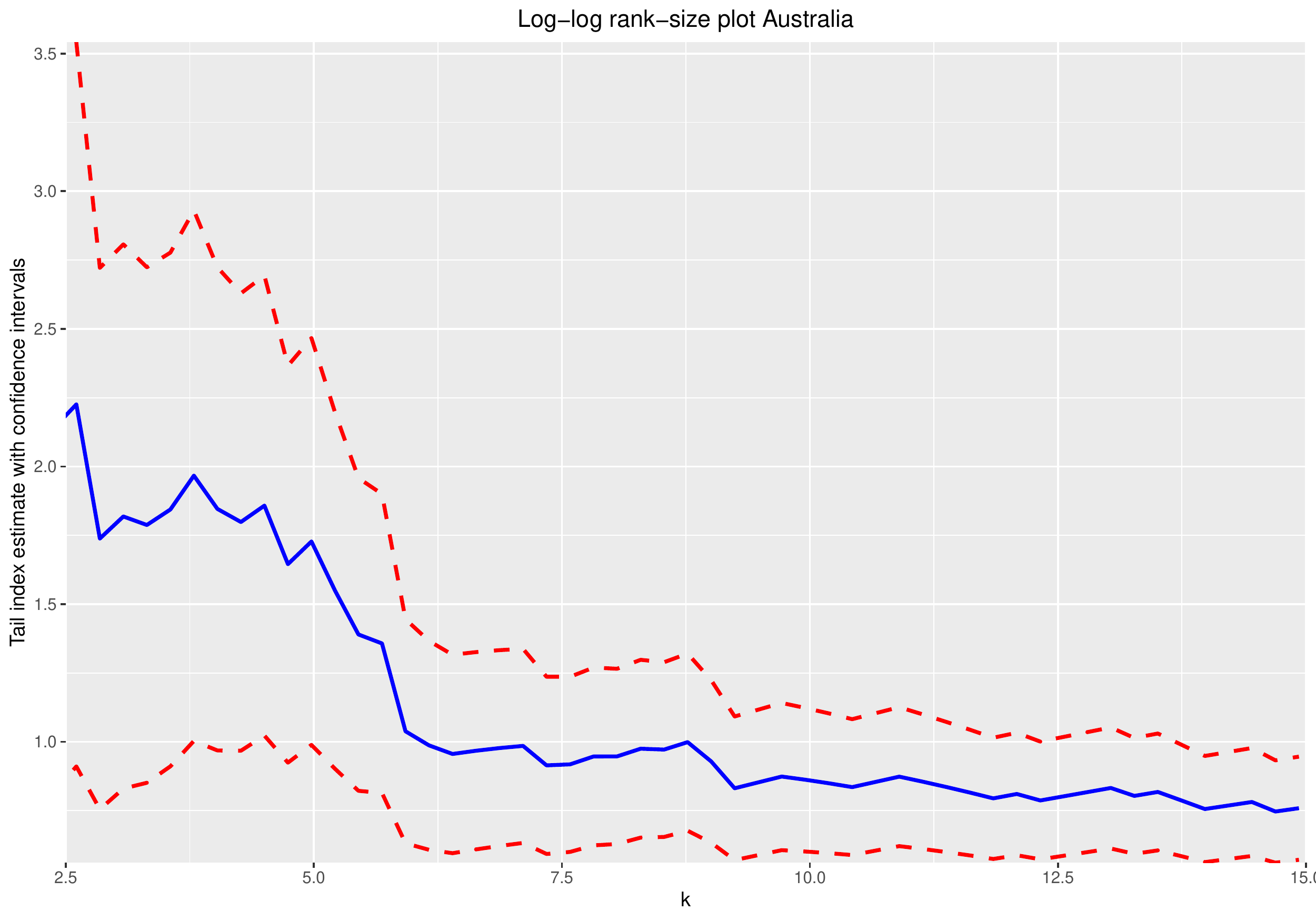}}
\subfigure[China]{\includegraphics[width=0.4\linewidth]{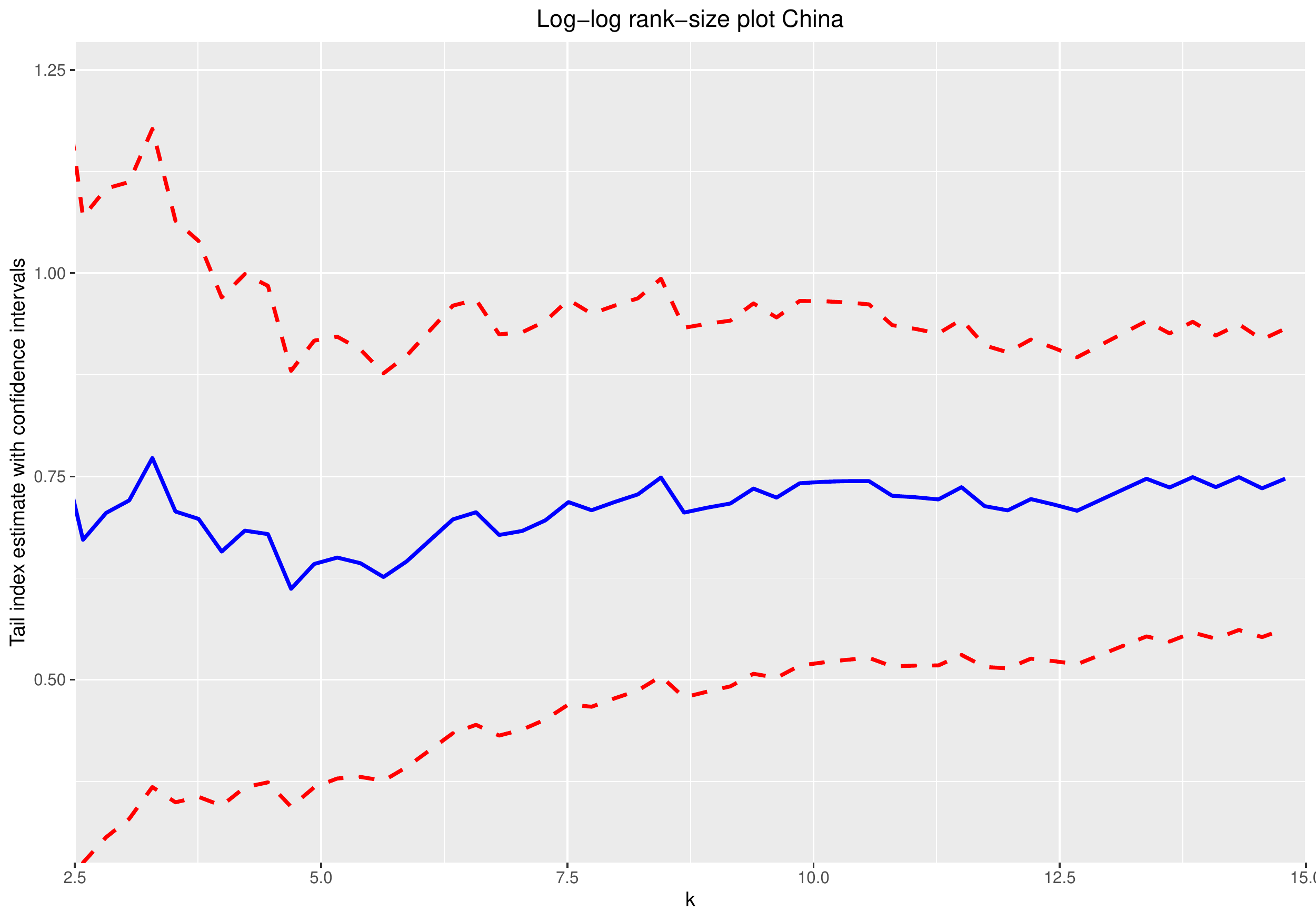}}\\
\subfigure[France]{\includegraphics[width=0.4\linewidth]{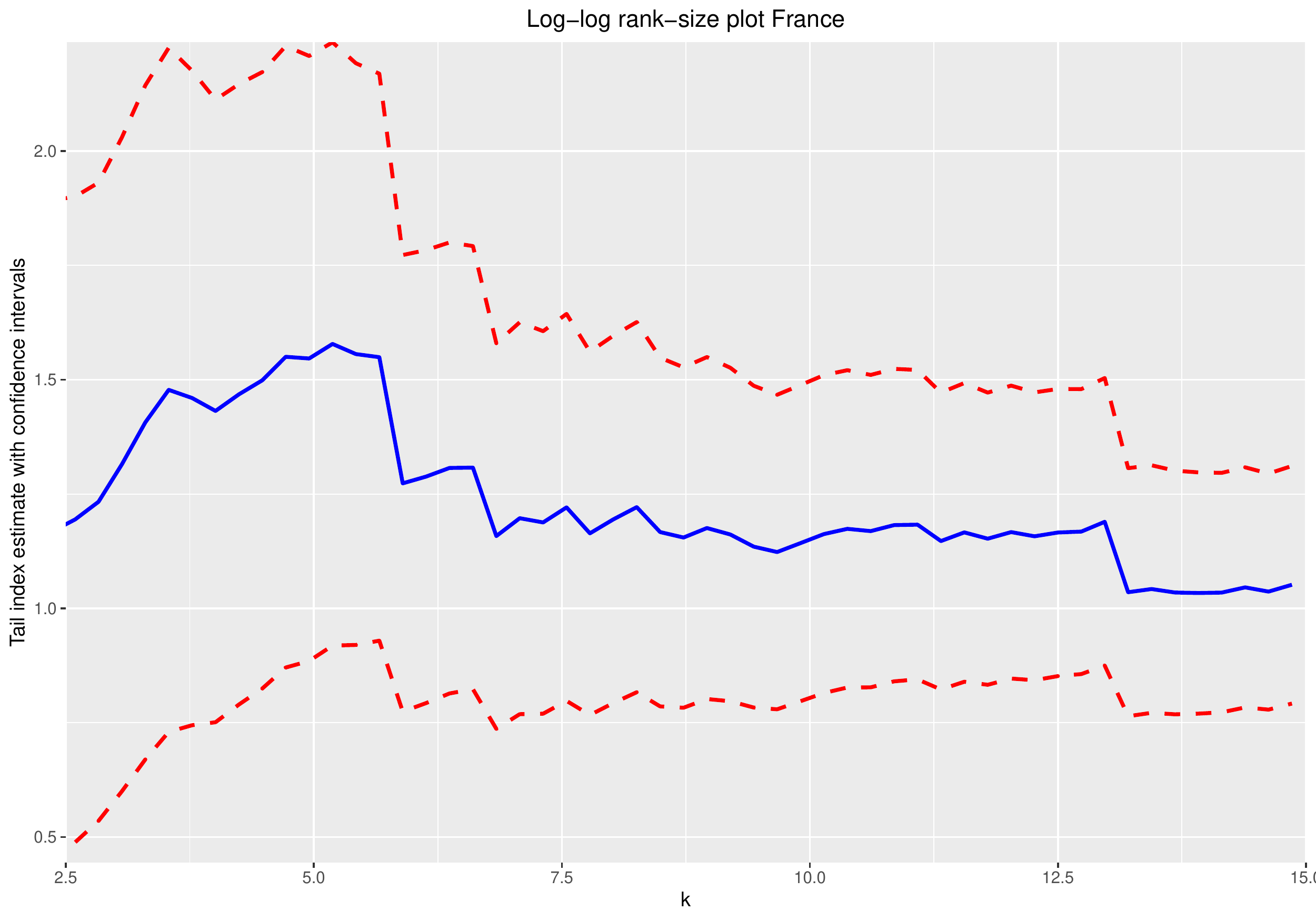}}
\subfigure[India]{\includegraphics[width=0.4\linewidth]{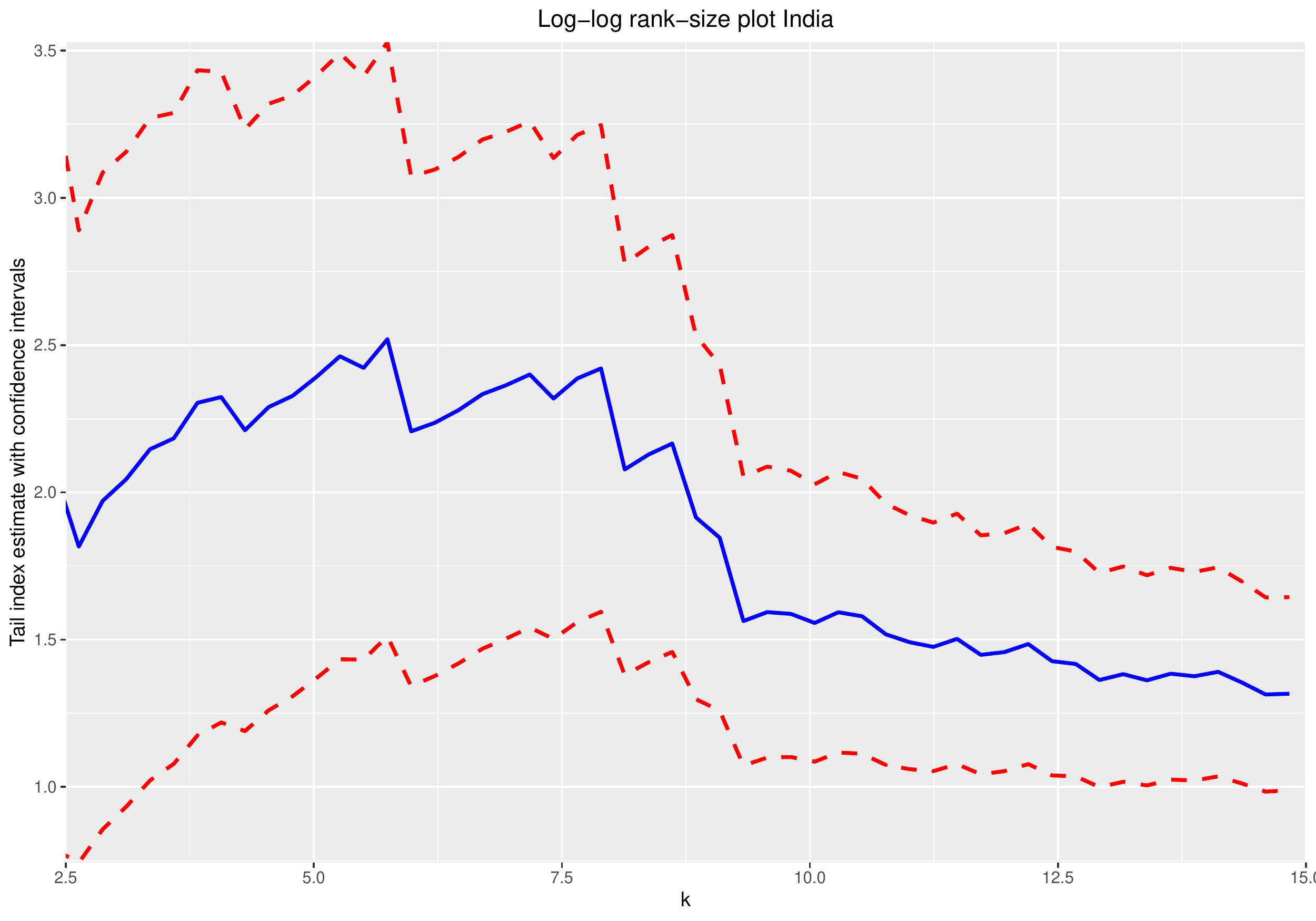}}
\end{center}%
\end{figure}

\setcounter{figure}{0}
\begin{figure}[h]%
\begin{center}%
\captcont{Hill's tail index estimates for positive changes in daily COVID-19 infections (ctd)}
\subfigure[Italy]{\includegraphics[width=0.4\linewidth]{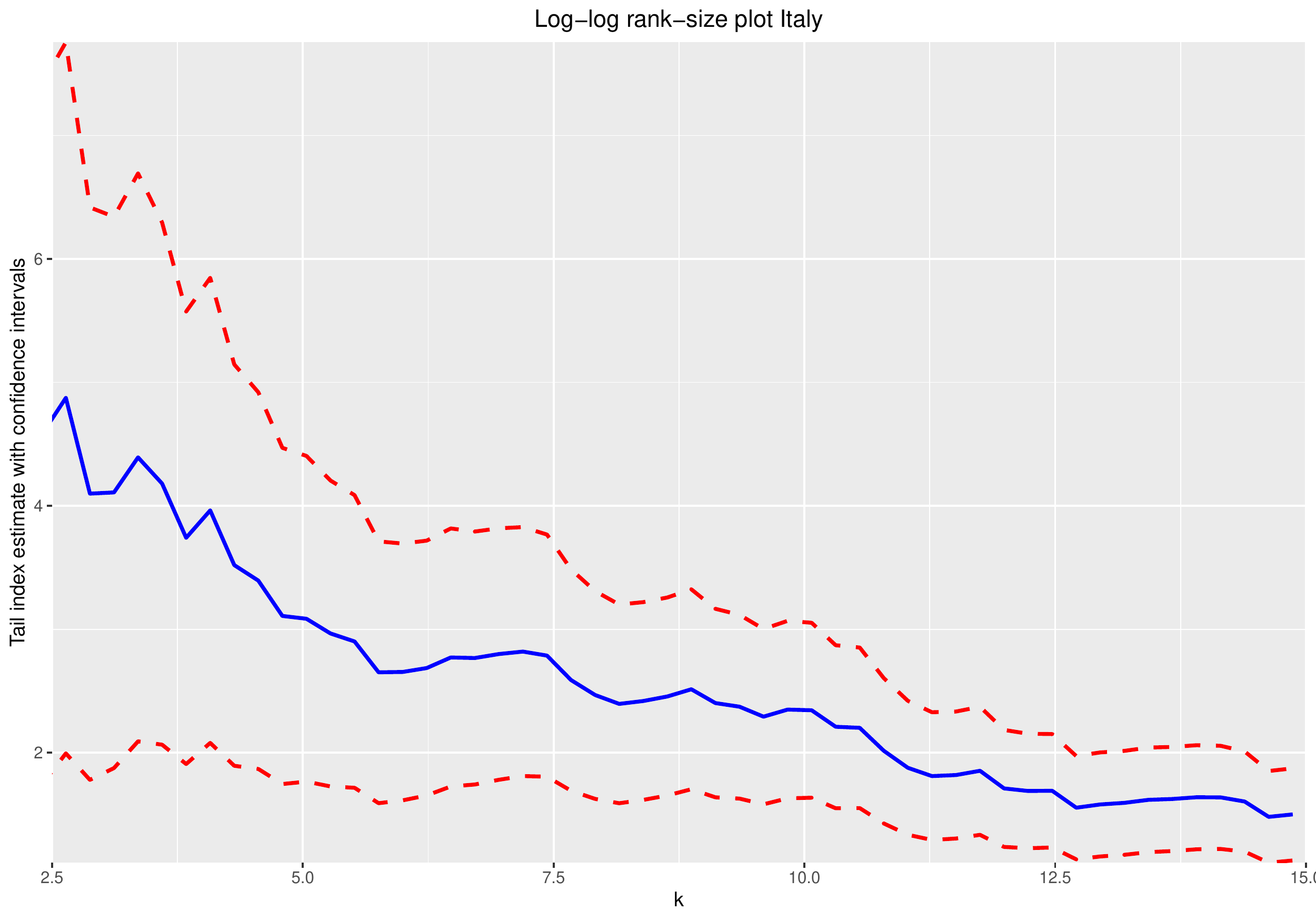}}
\subfigure[Russia]{\includegraphics[width=0.4\linewidth]{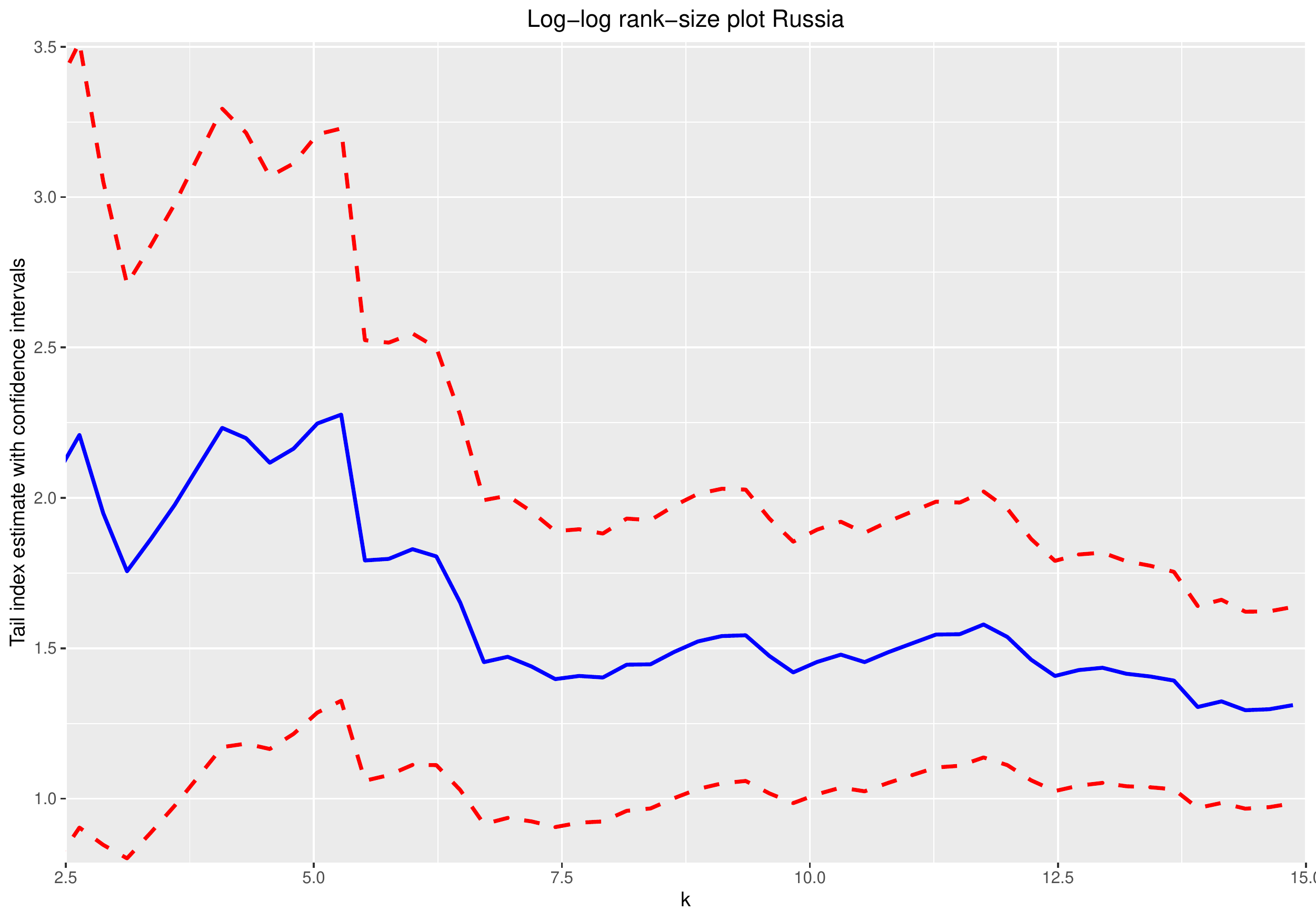}}\\
\subfigure[UK]{\includegraphics[width=0.4\linewidth]{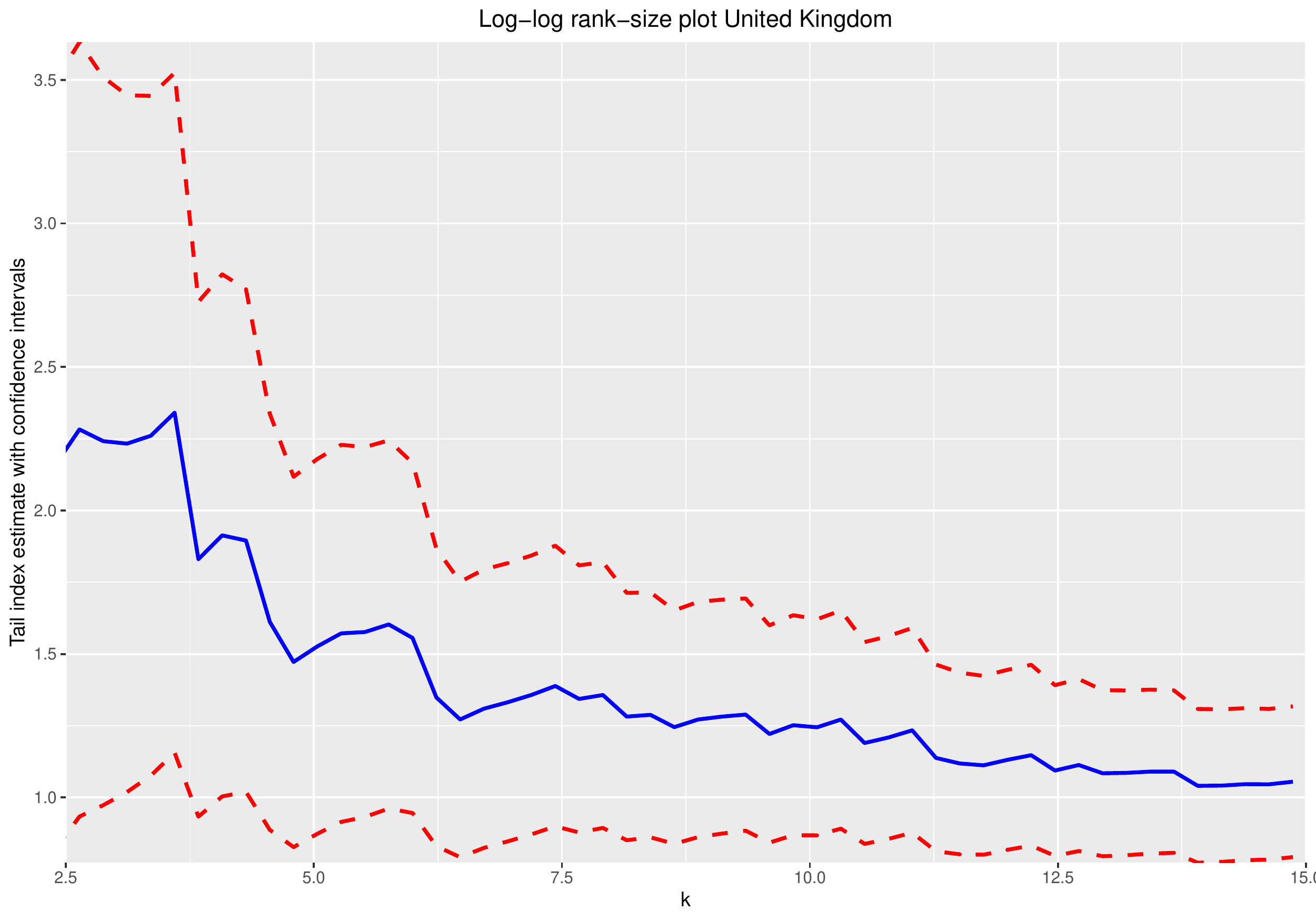}}
\subfigure[US]{\includegraphics[width=0.4\linewidth]{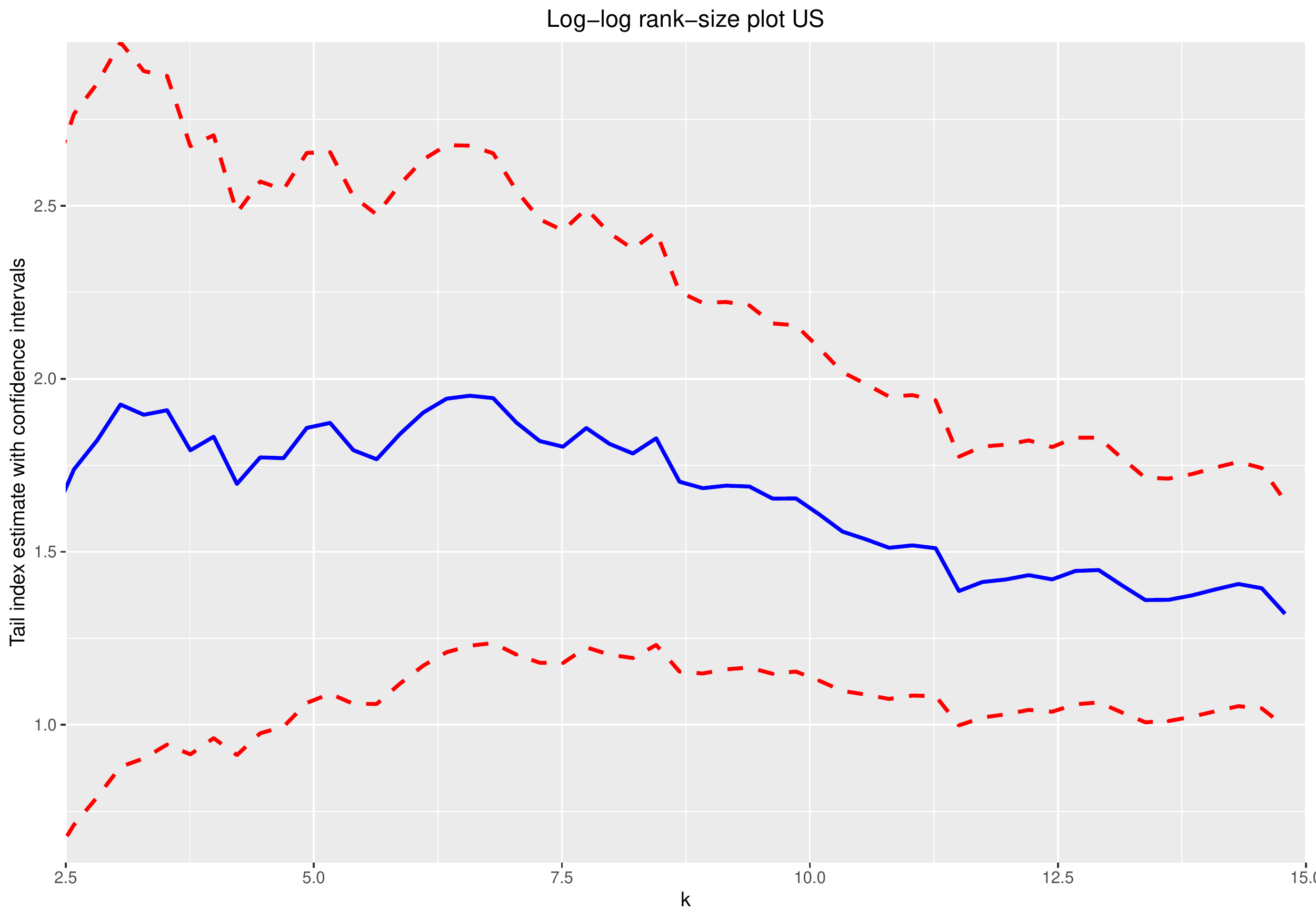}}
\end{center}
\end{figure}

\setcounter{figure}{1}

\begin{figure}[!h]%
\caption{Hill's tail index estimates for positive changes in daily COVID-19 deaths}\label{Fig2}
\begin{center}%
\subfigure[Australia]{\includegraphics[width=0.4\linewidth]{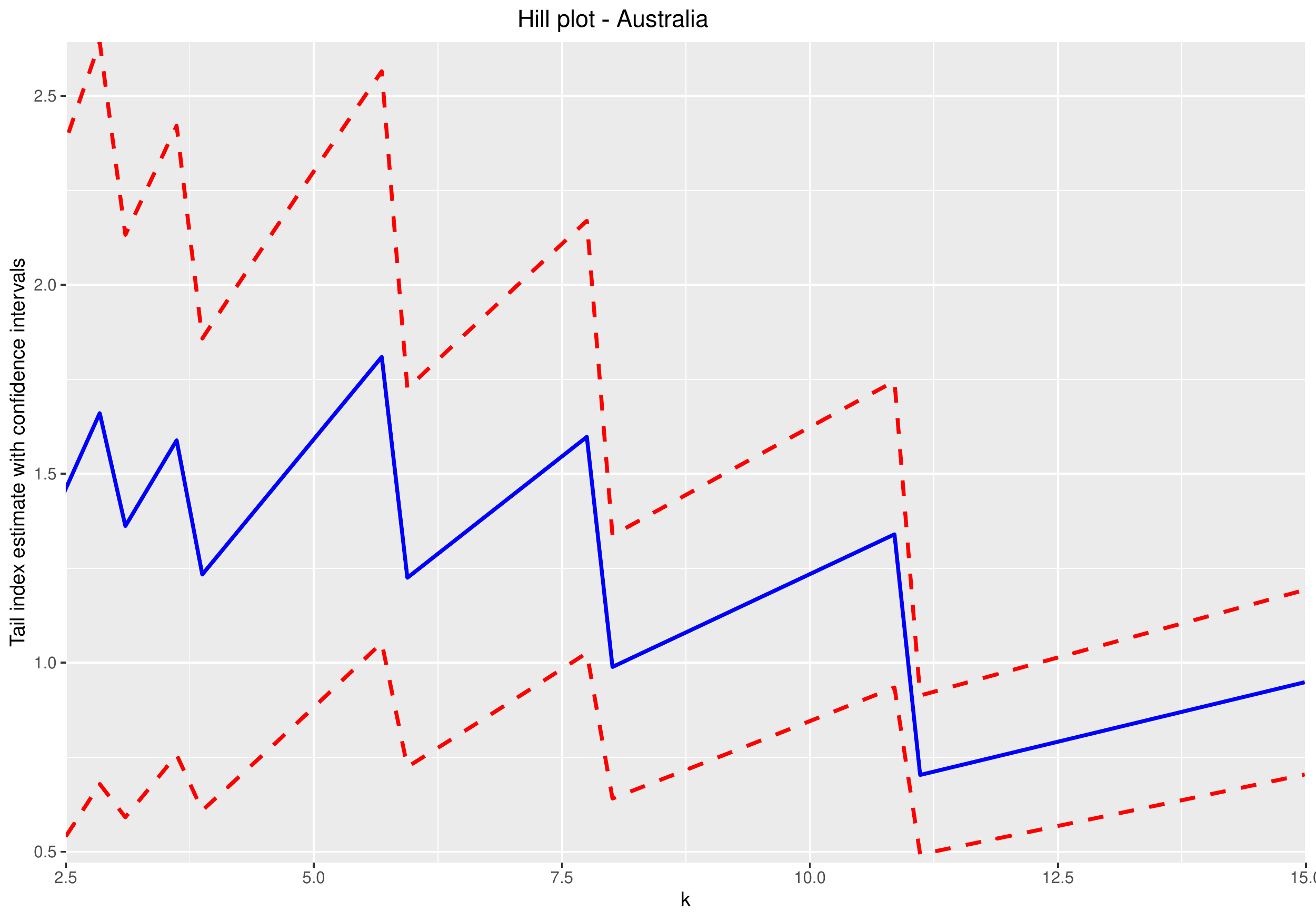}}
\subfigure[China]{\includegraphics[width=0.4\linewidth]{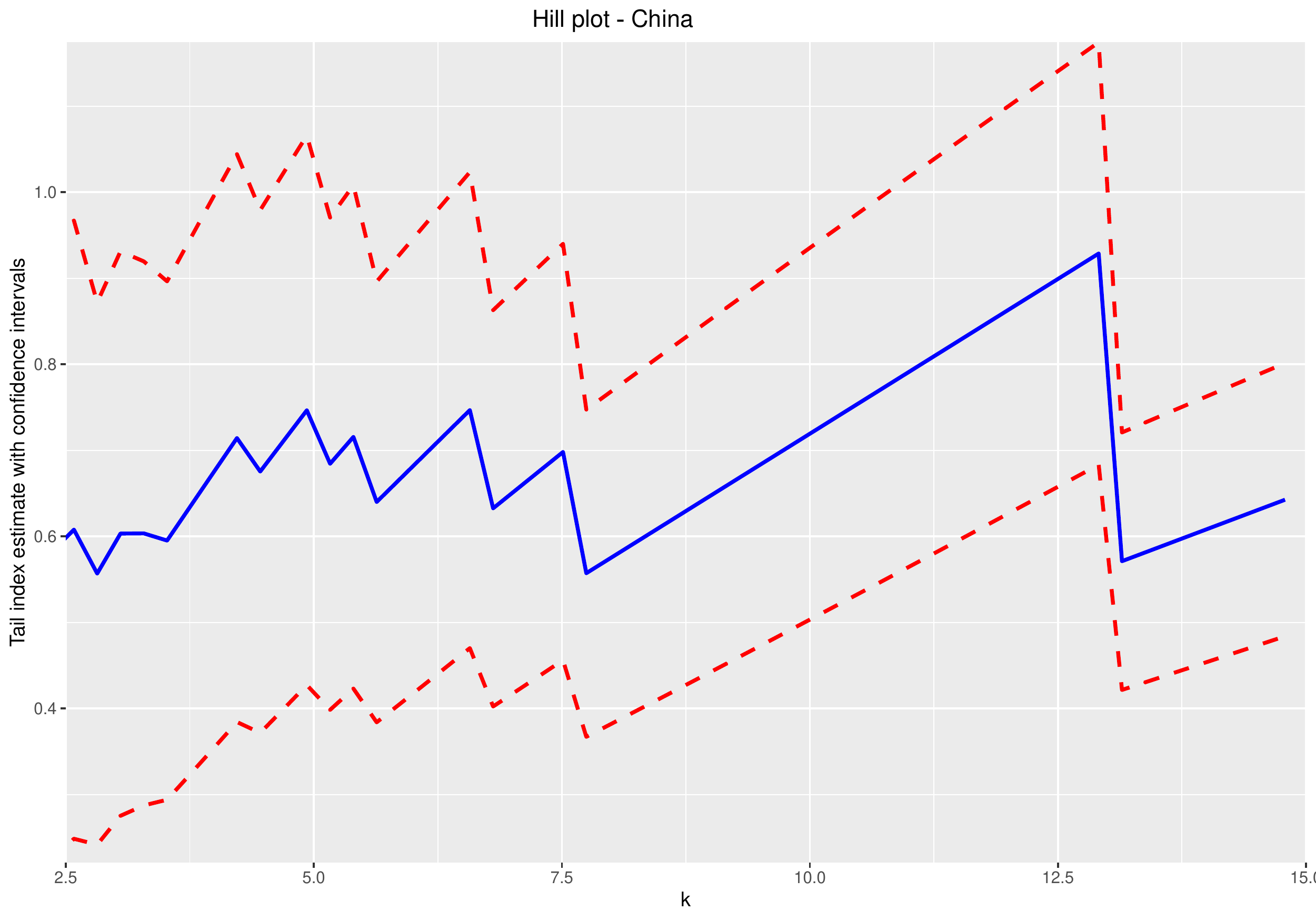}}\\
\subfigure[France]{\includegraphics[width=0.4\linewidth]{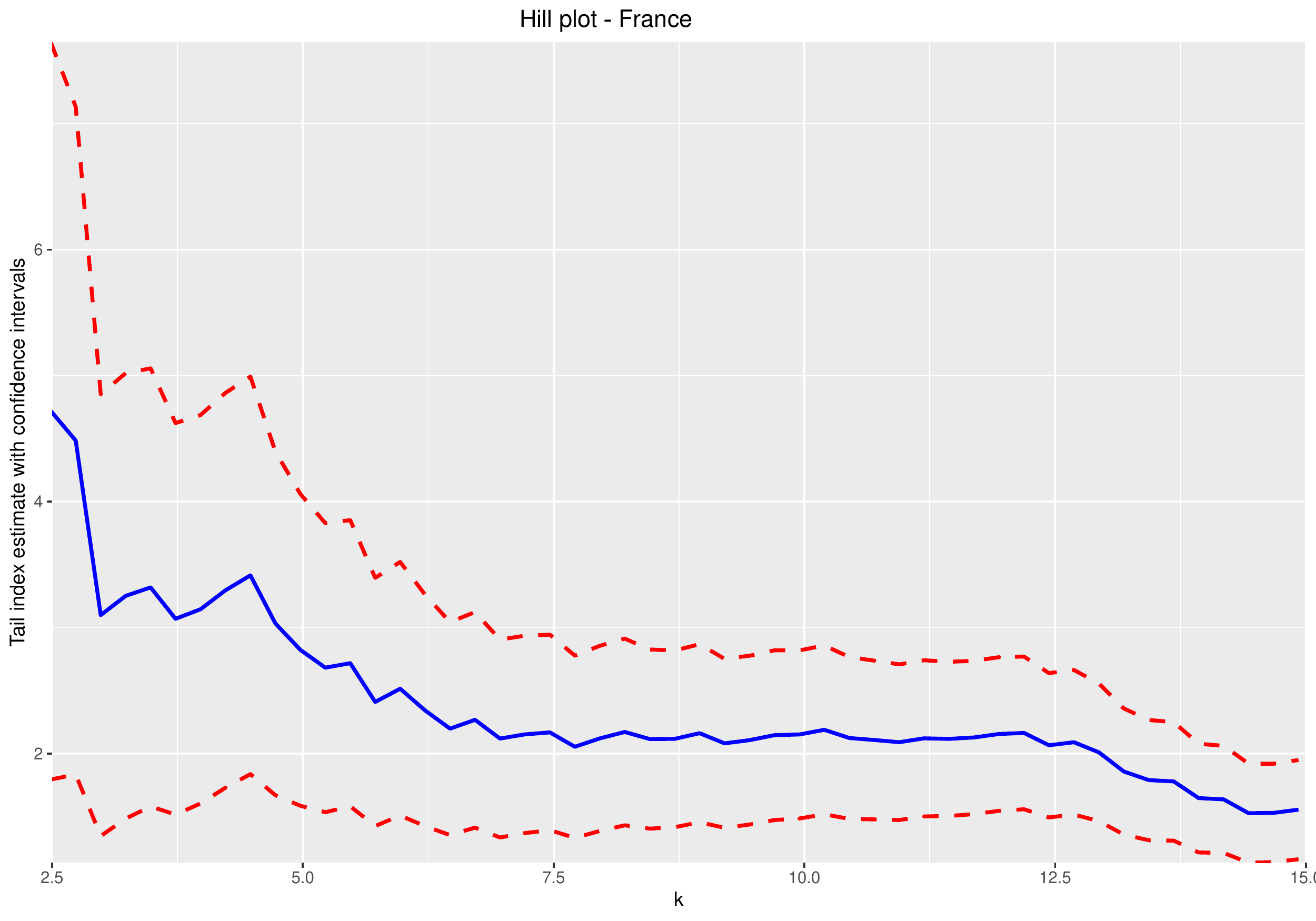}}
\subfigure[India]{\includegraphics[width=0.4\linewidth]{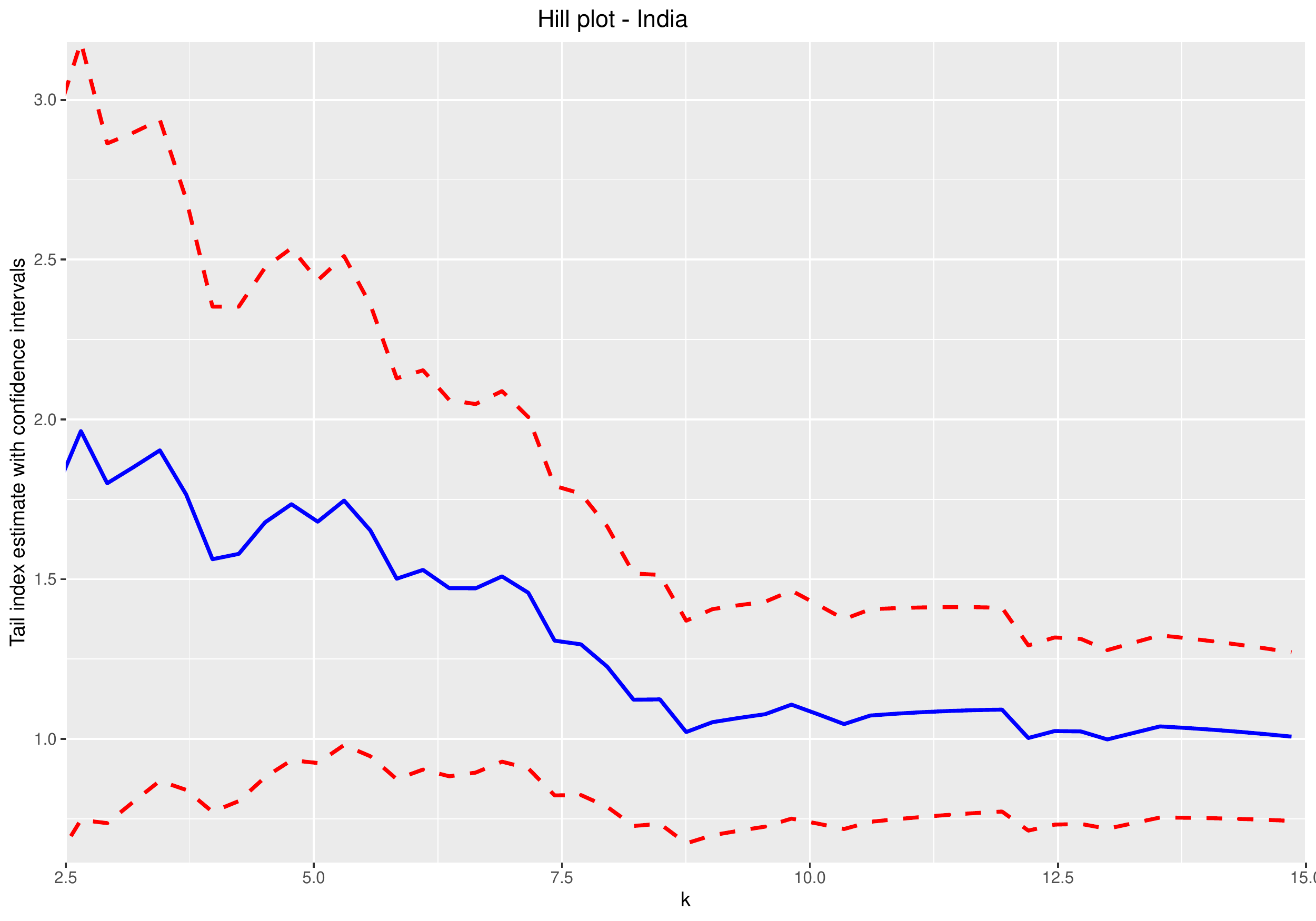}}
\end{center}%
\end{figure}

\setcounter{figure}{1}
\begin{figure}[h]%
\begin{center}%
\captcont{Hill's tail index estimates for positive changes in daily COVID-19 deaths (ctd)}
\subfigure[Italy]{\includegraphics[width=0.4\linewidth]{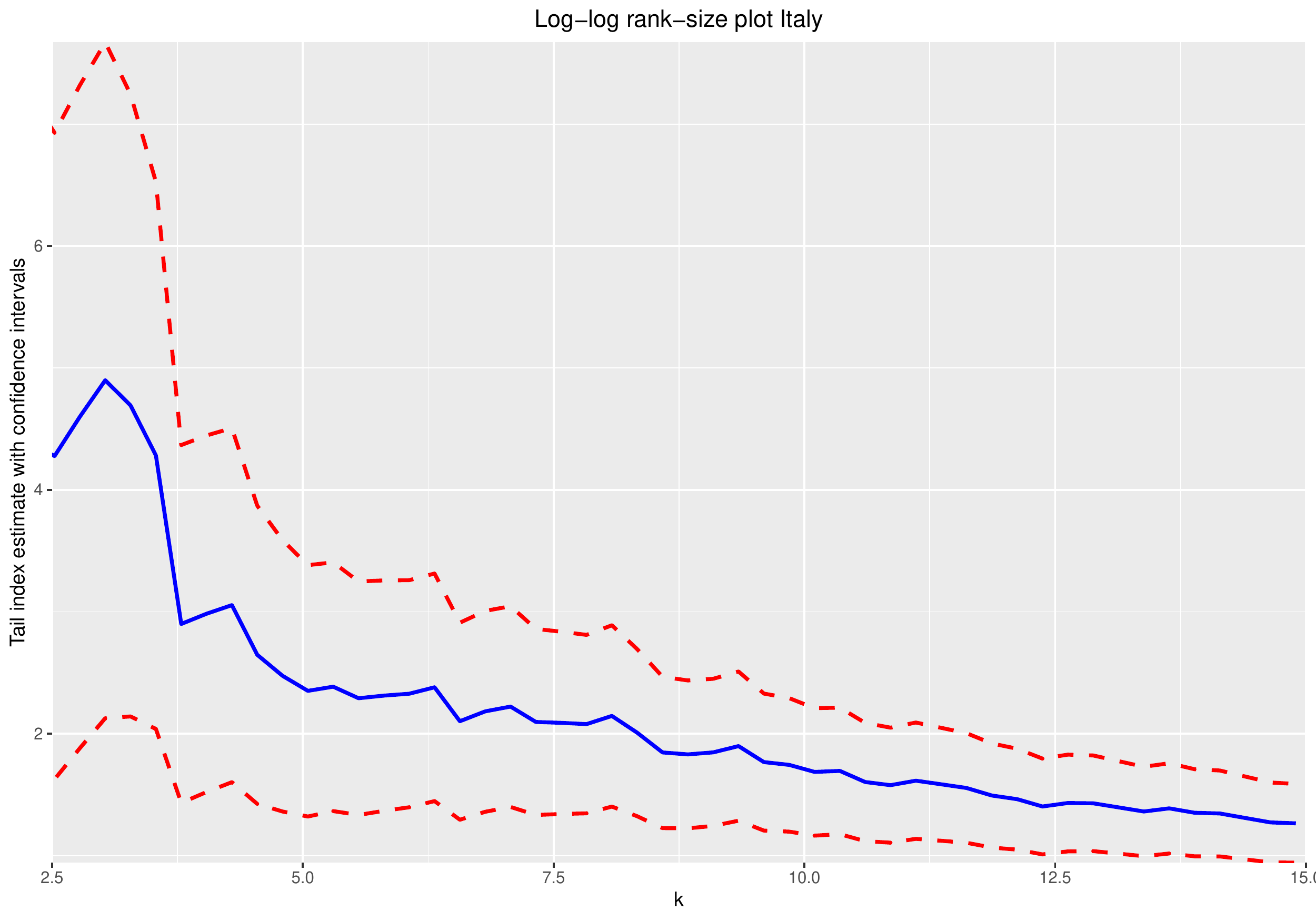}}
\subfigure[Russia]{\includegraphics[width=0.4\linewidth]{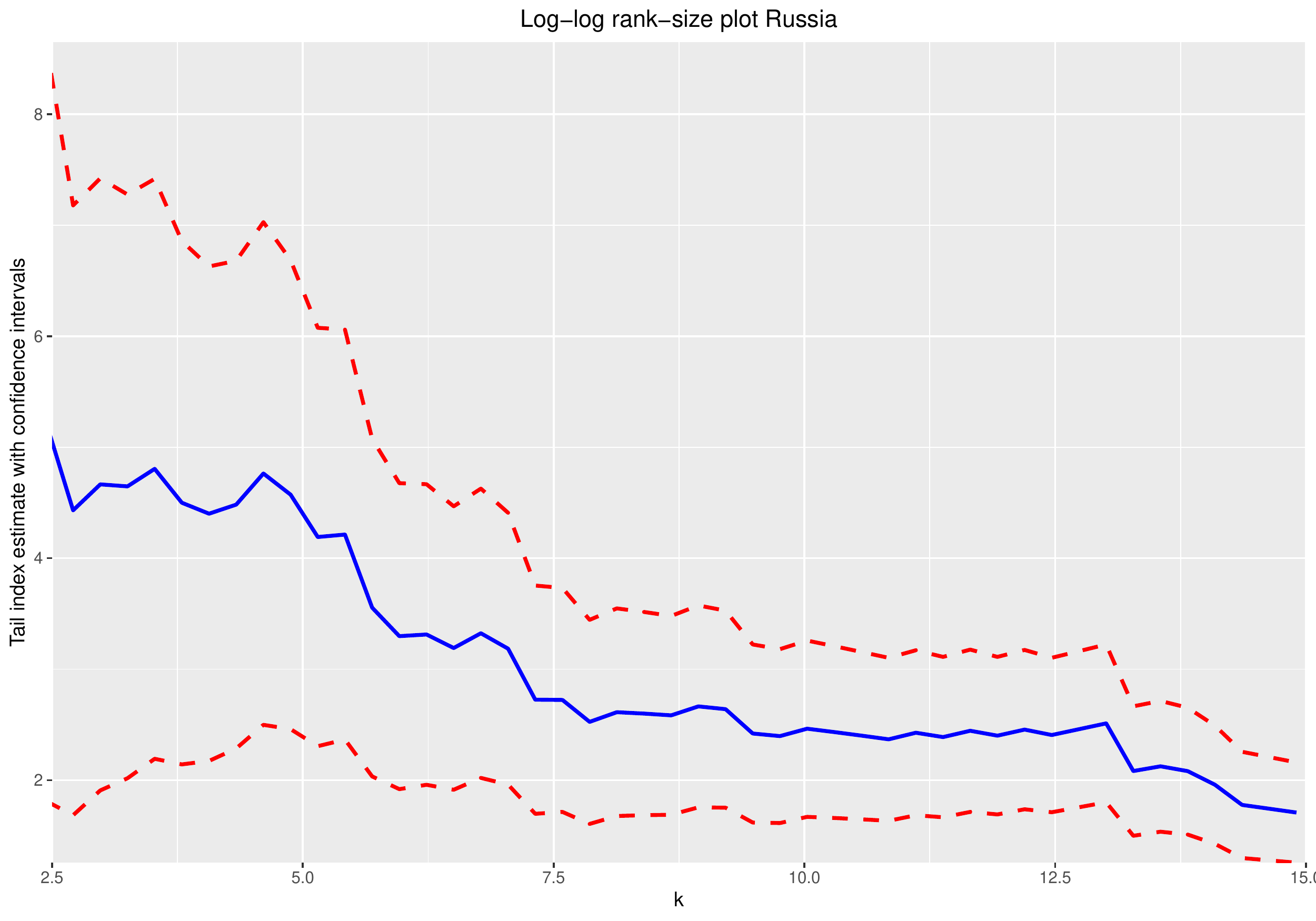}}\\
\subfigure[UK]{\includegraphics[width=0.4\linewidth]{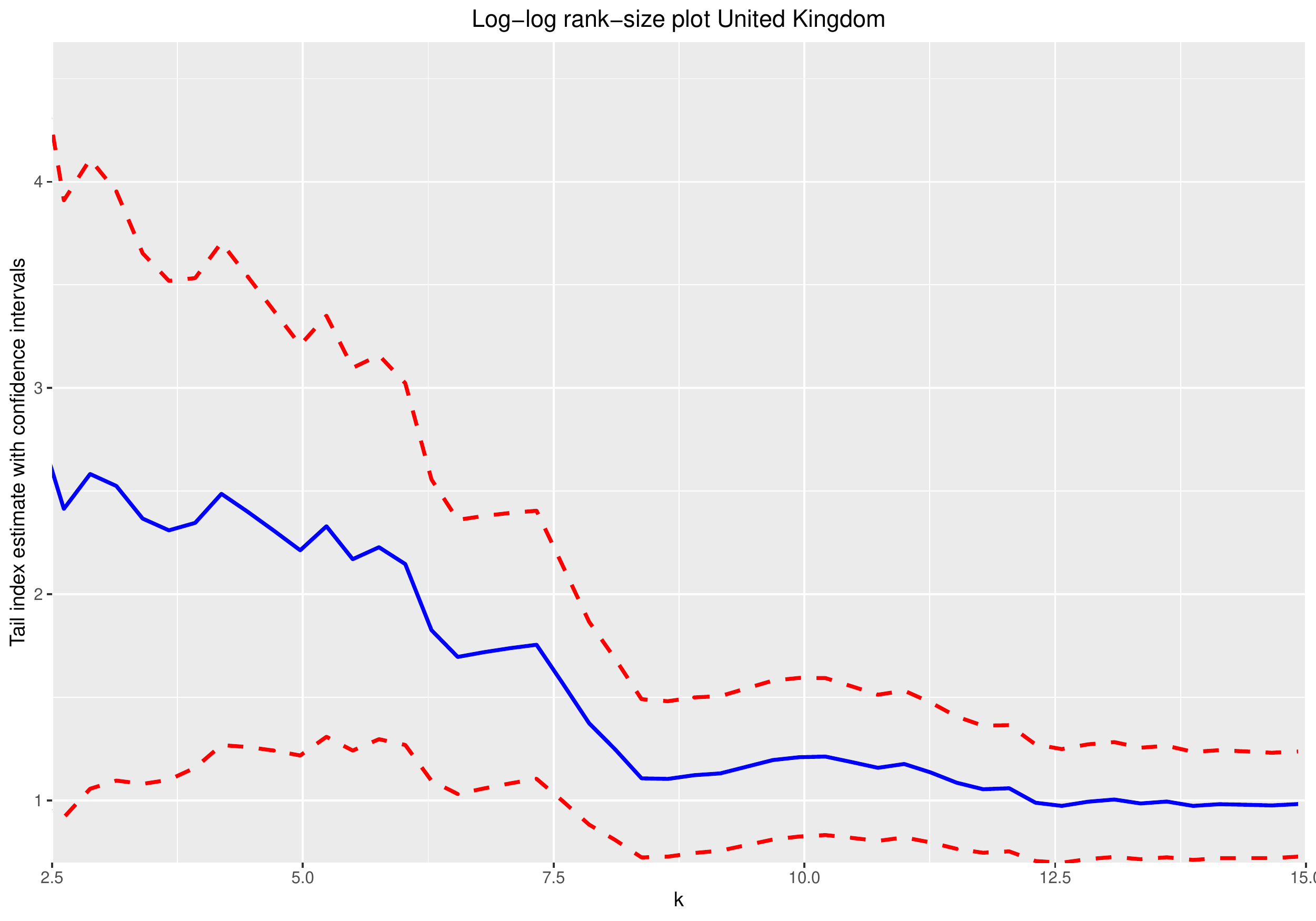}}
\subfigure[US]{\includegraphics[width=0.4\linewidth]{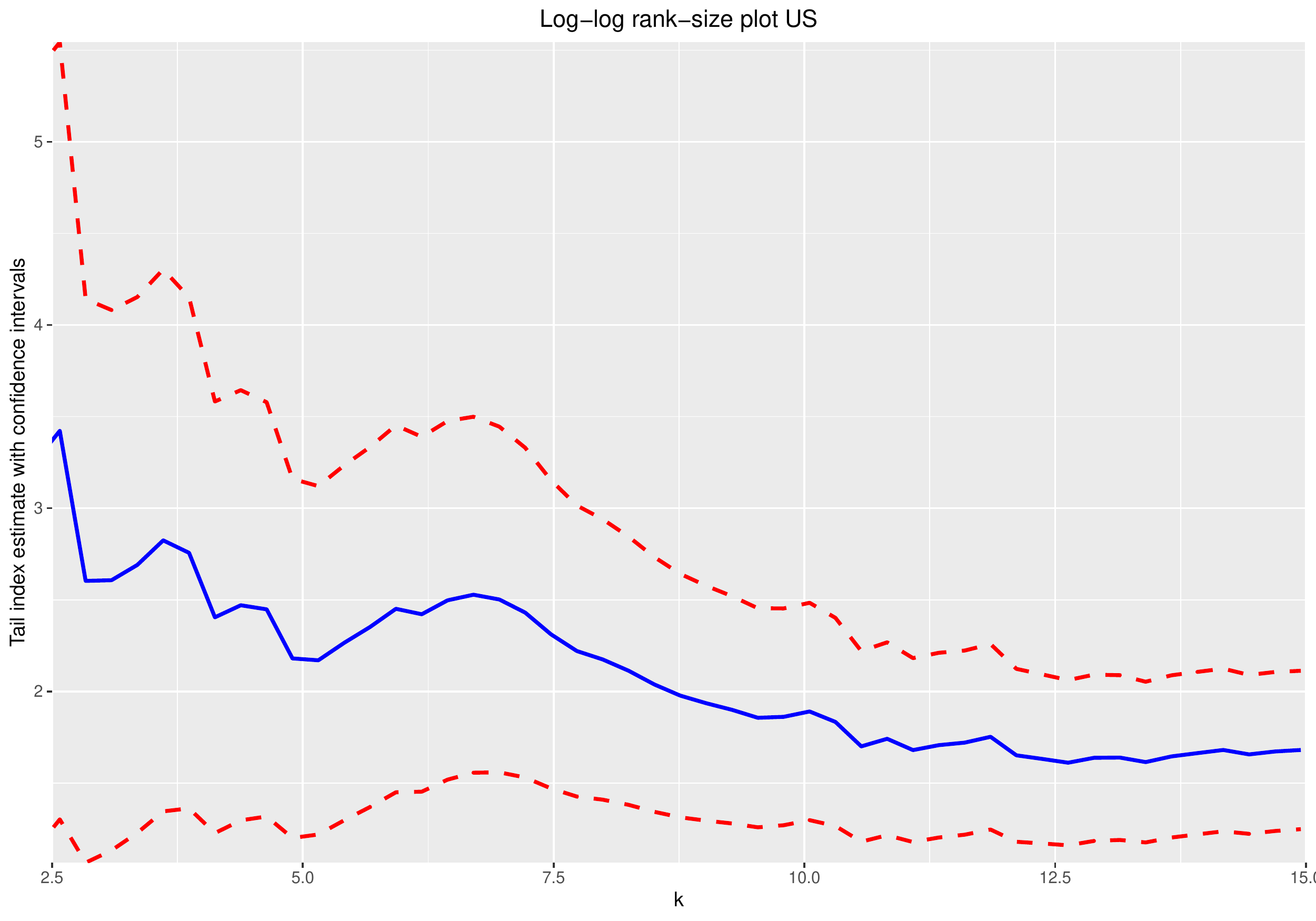}}
\end{center}
\end{figure}

\setcounter{figure}{1}

\setcounter{figure}{2}
\begin{figure}[h]%
\caption{Log-log rank-size regression tail index estimates for positive changes in daily COVID-19 infections}\label{Fig3}
\begin{center}%
\subfigure[Australia]{\includegraphics[width=0.4\linewidth]{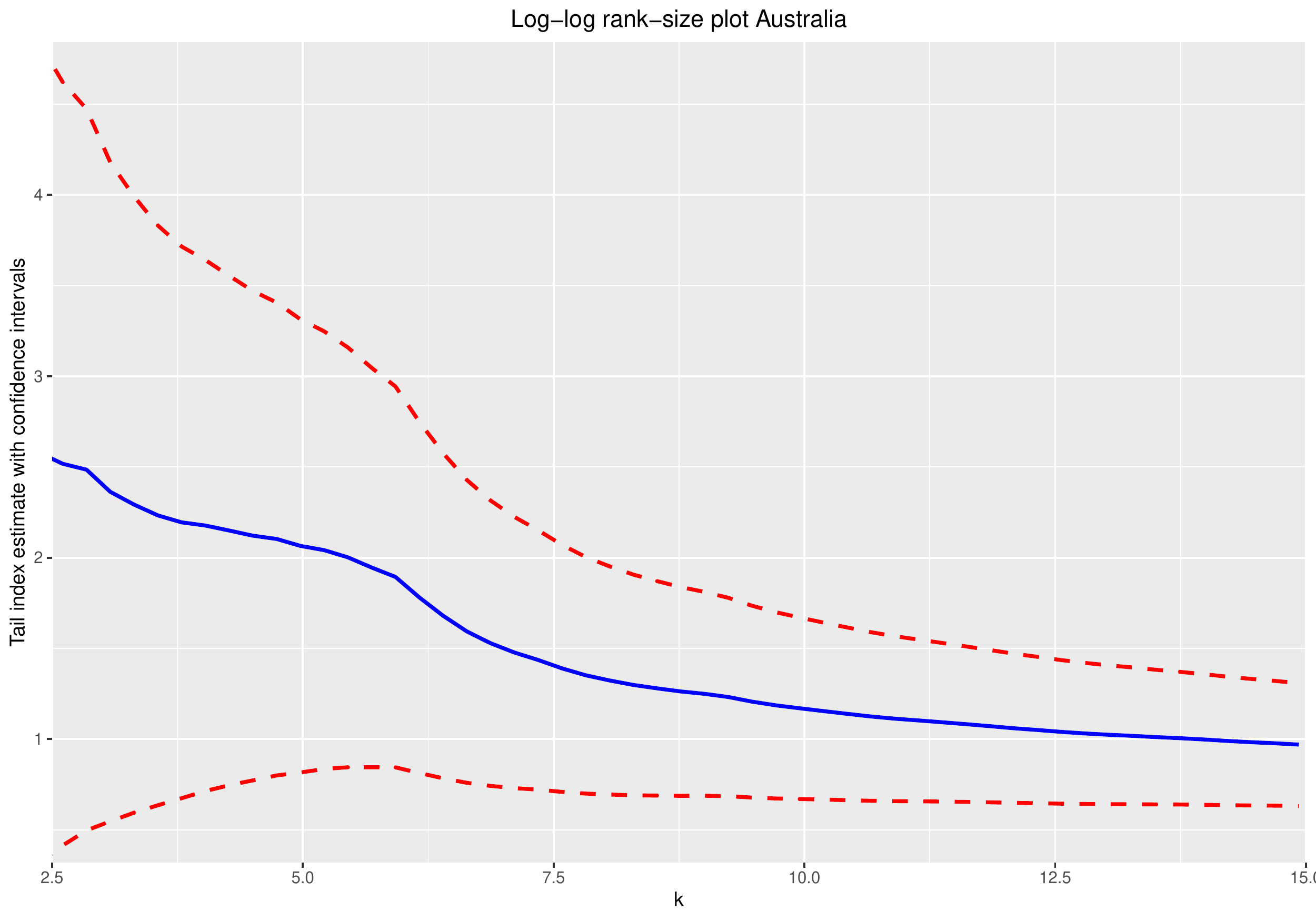}}
\subfigure[China]{\includegraphics[width=0.4\linewidth]{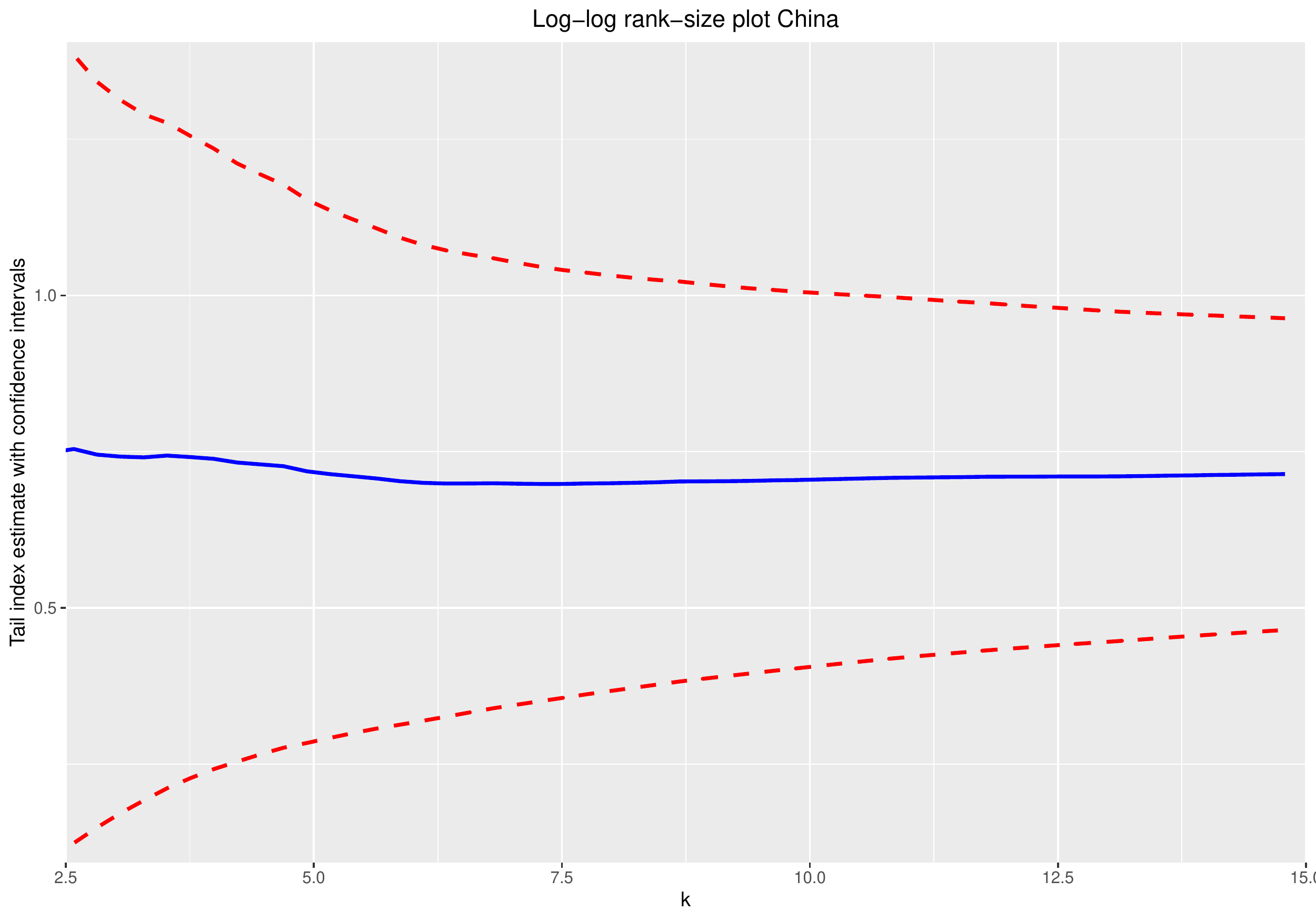}}\\
\subfigure[France]{\includegraphics[width=0.4\linewidth]{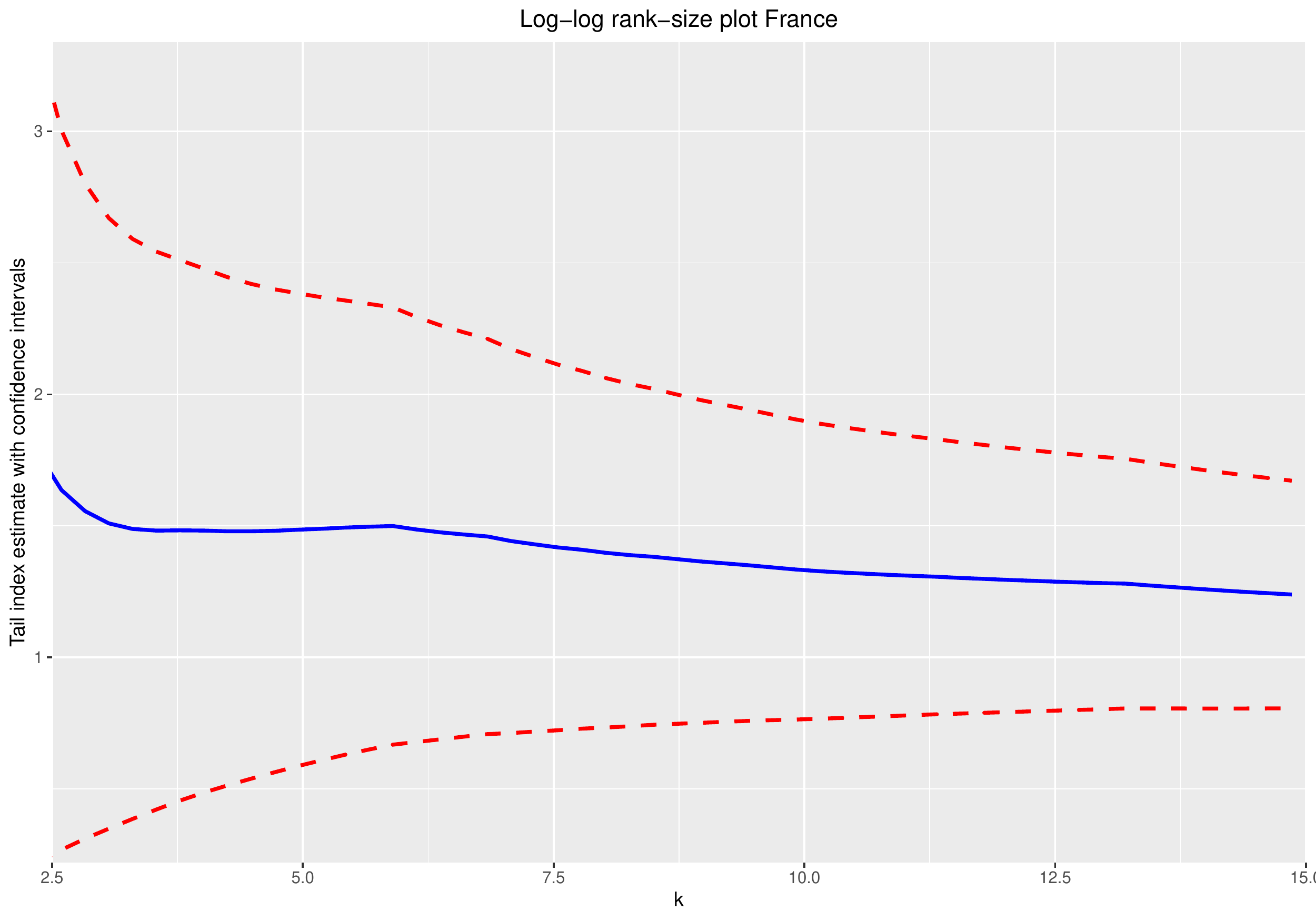}}
\subfigure[India]{\includegraphics[width=0.4\linewidth]{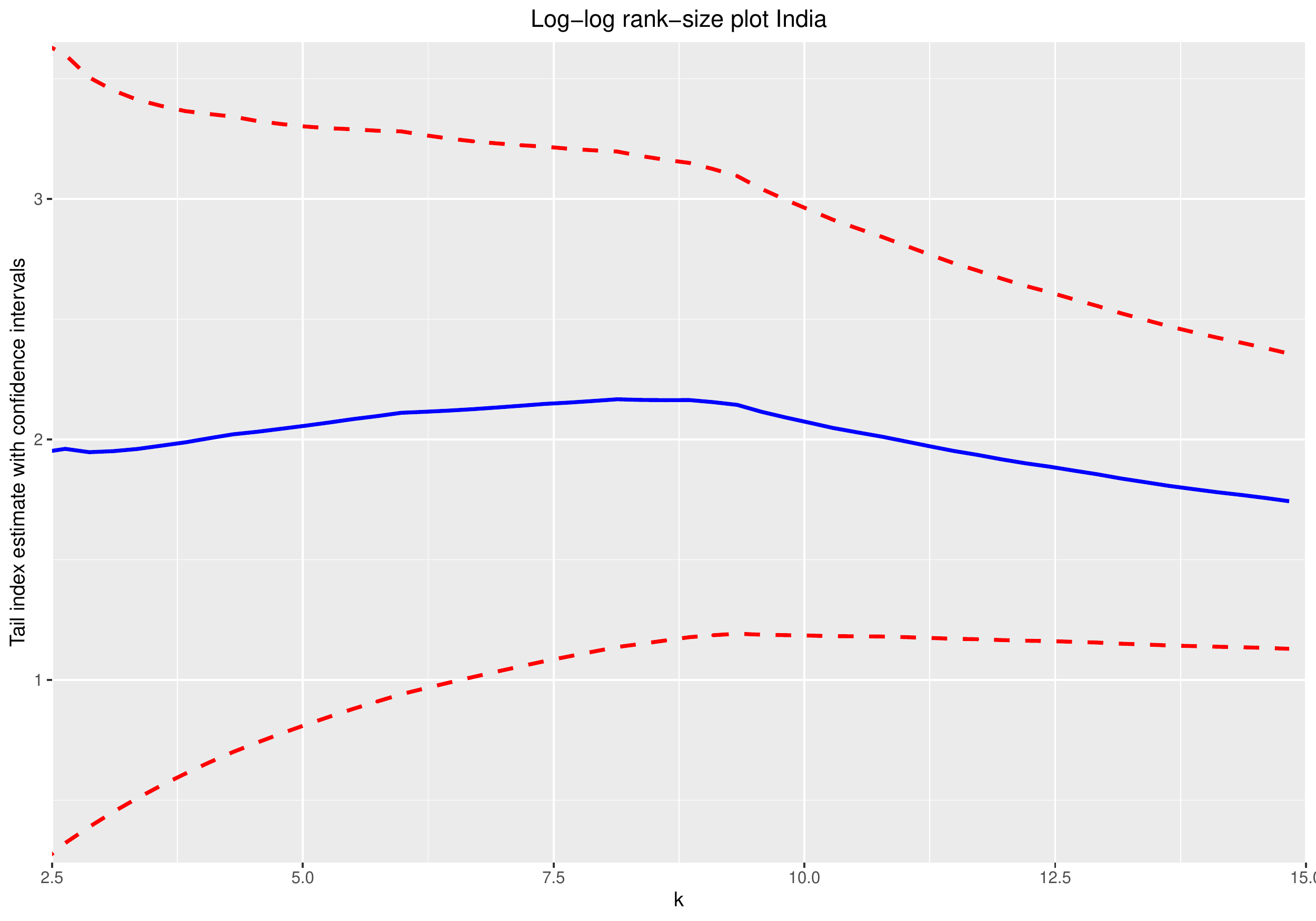}}
\end{center}%
\end{figure}

\setcounter{figure}{2}
\begin{figure}[h]%
\begin{center}%
\captcont{Log-log rank-size regression tail index estimates for positive changes in daily COVID-19 infections (ctd)}
\subfigure[Italy]{\includegraphics[width=0.4\linewidth]{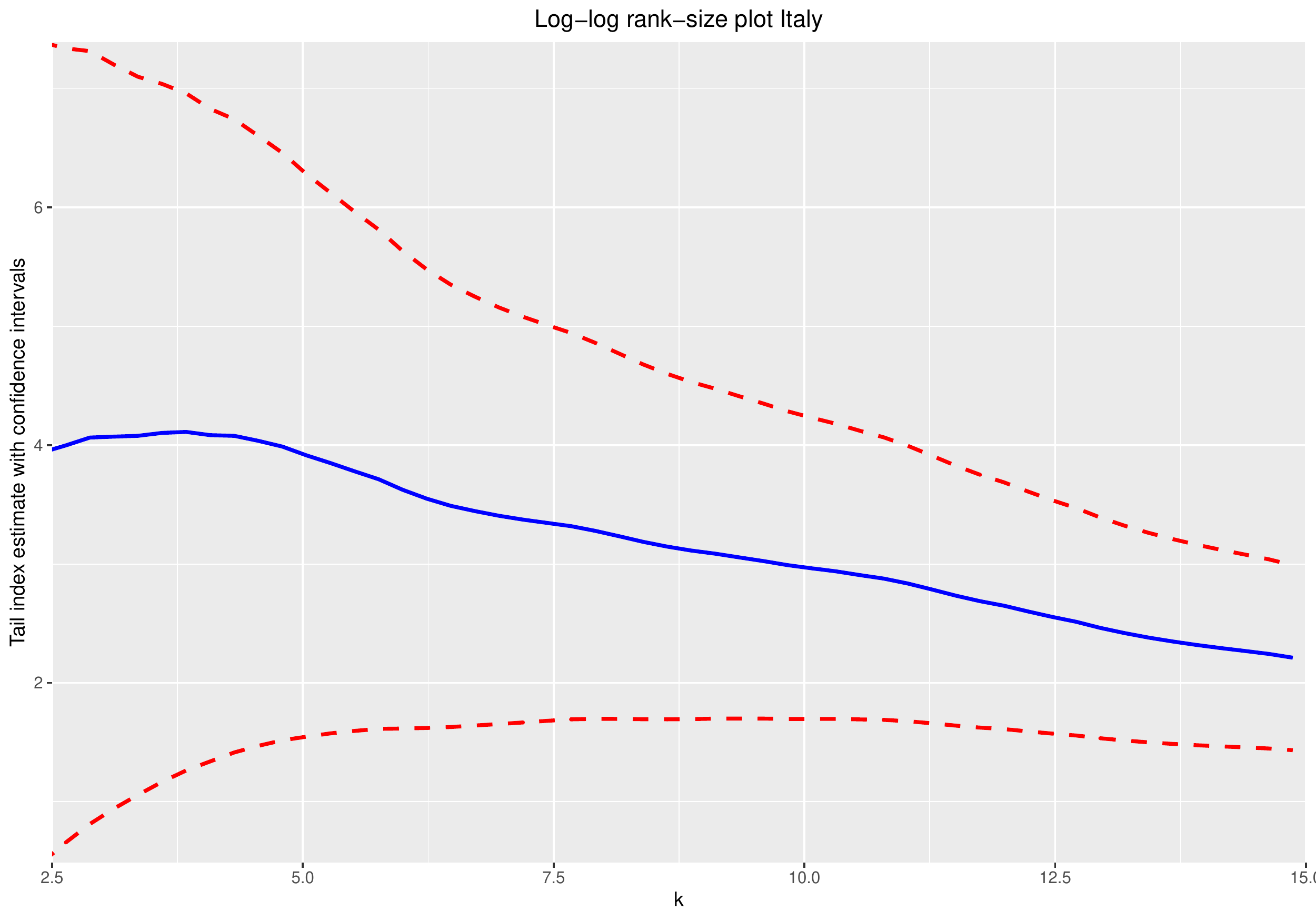}}
\subfigure[Russia]{\includegraphics[width=0.4\linewidth]{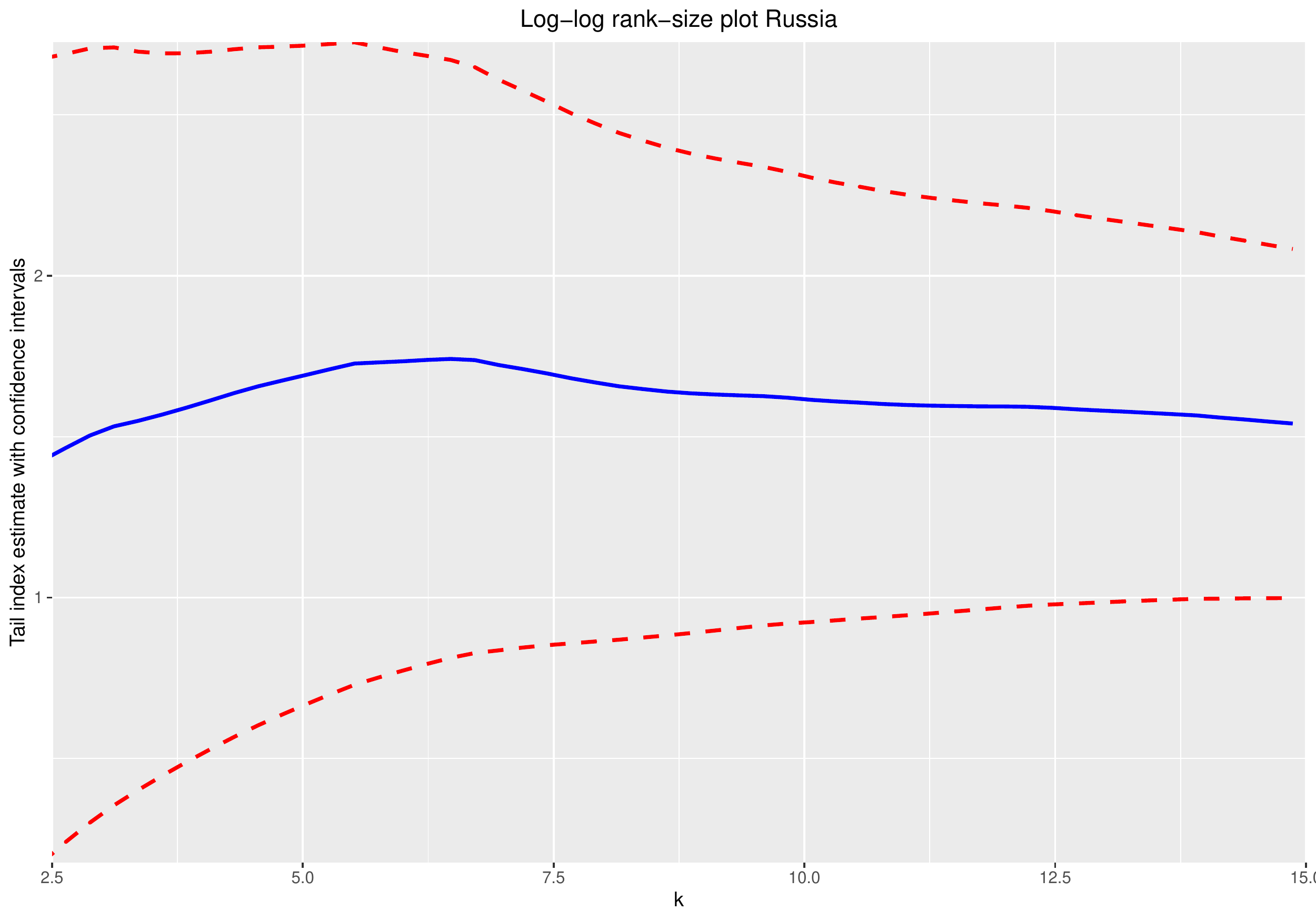}}\\
\subfigure[UK]{\includegraphics[width=0.4\linewidth]{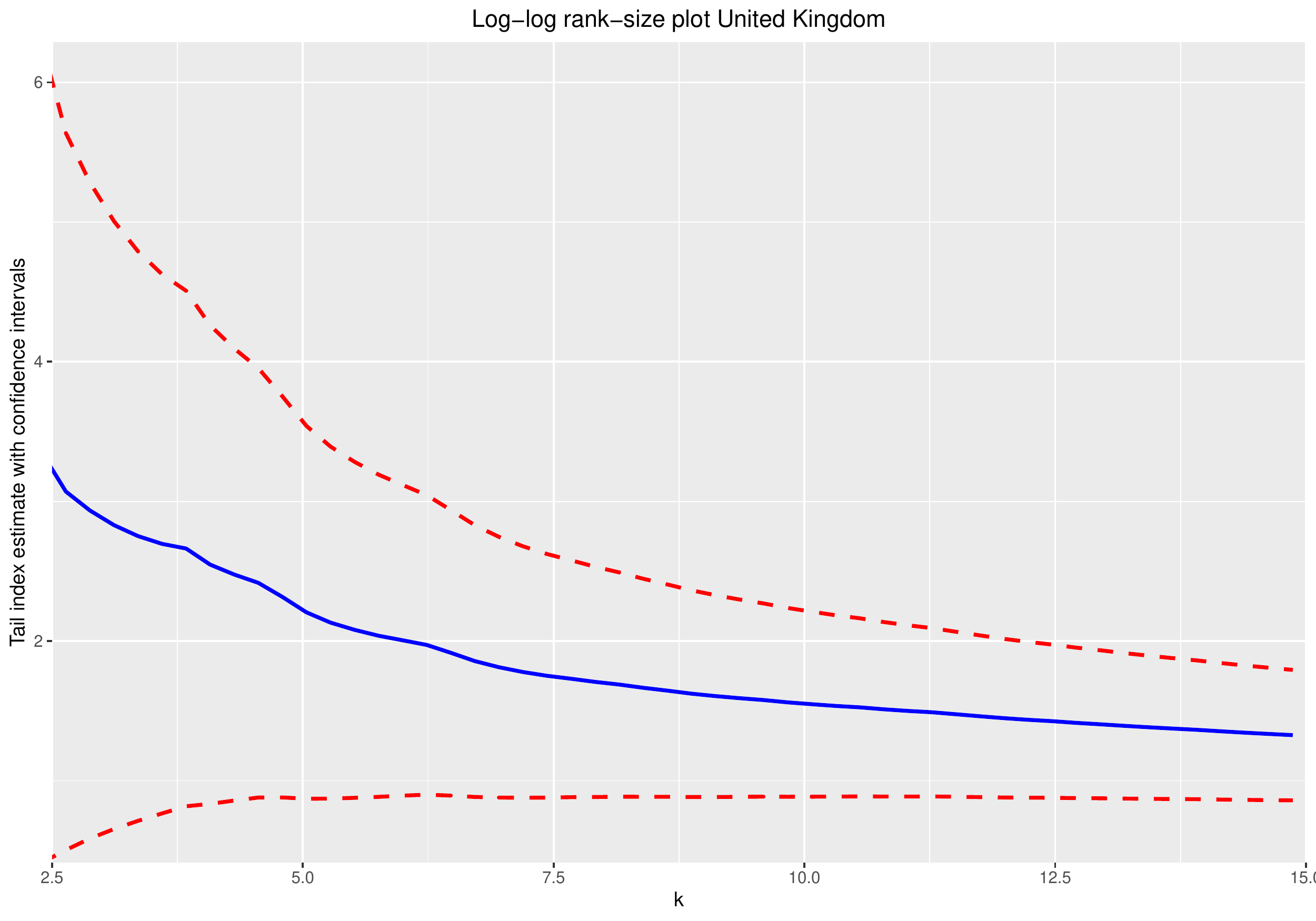}}
\subfigure[US]{\includegraphics[width=0.4\linewidth]{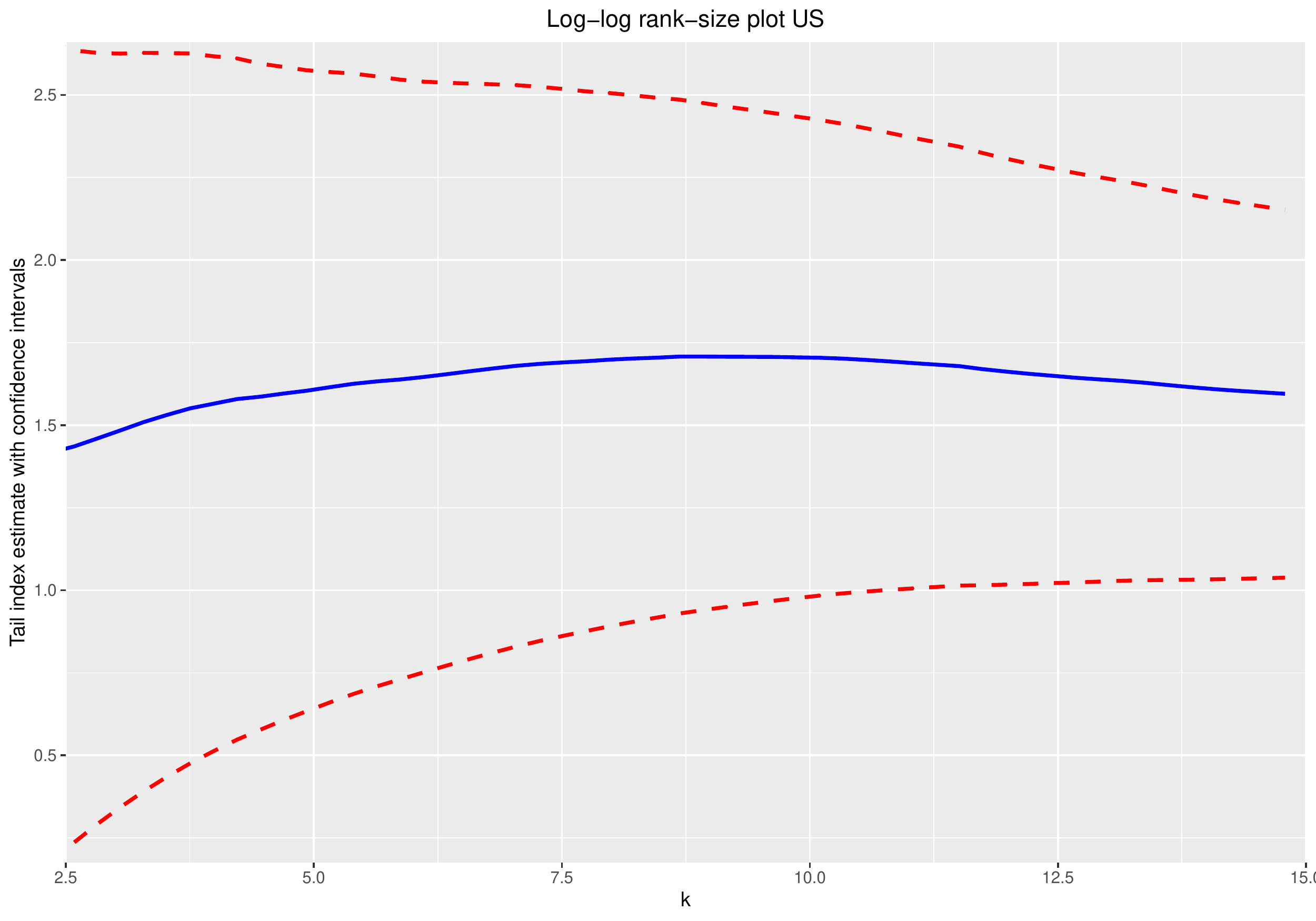}}
\end{center}
\end{figure}

\setcounter{figure}{3}
\begin{figure}[h]%
\caption{Log-log rank-size regression tail index estimates for positive changes in daily COVID-19 deaths}\label{Fig4}
\begin{center}%
\subfigure[Australia]{\includegraphics[width=0.4\linewidth]{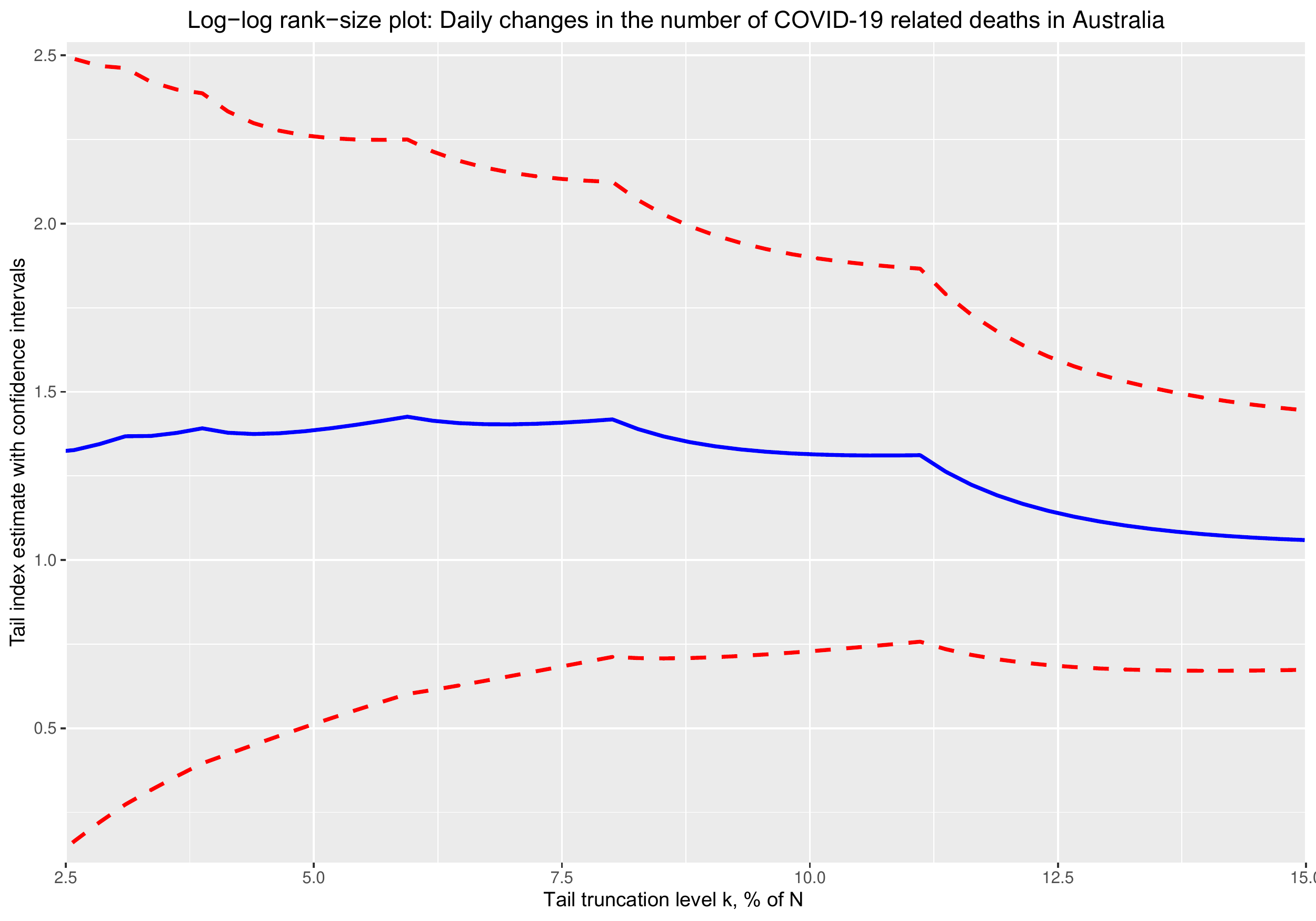}}
\subfigure[China]{\includegraphics[width=0.4\linewidth]{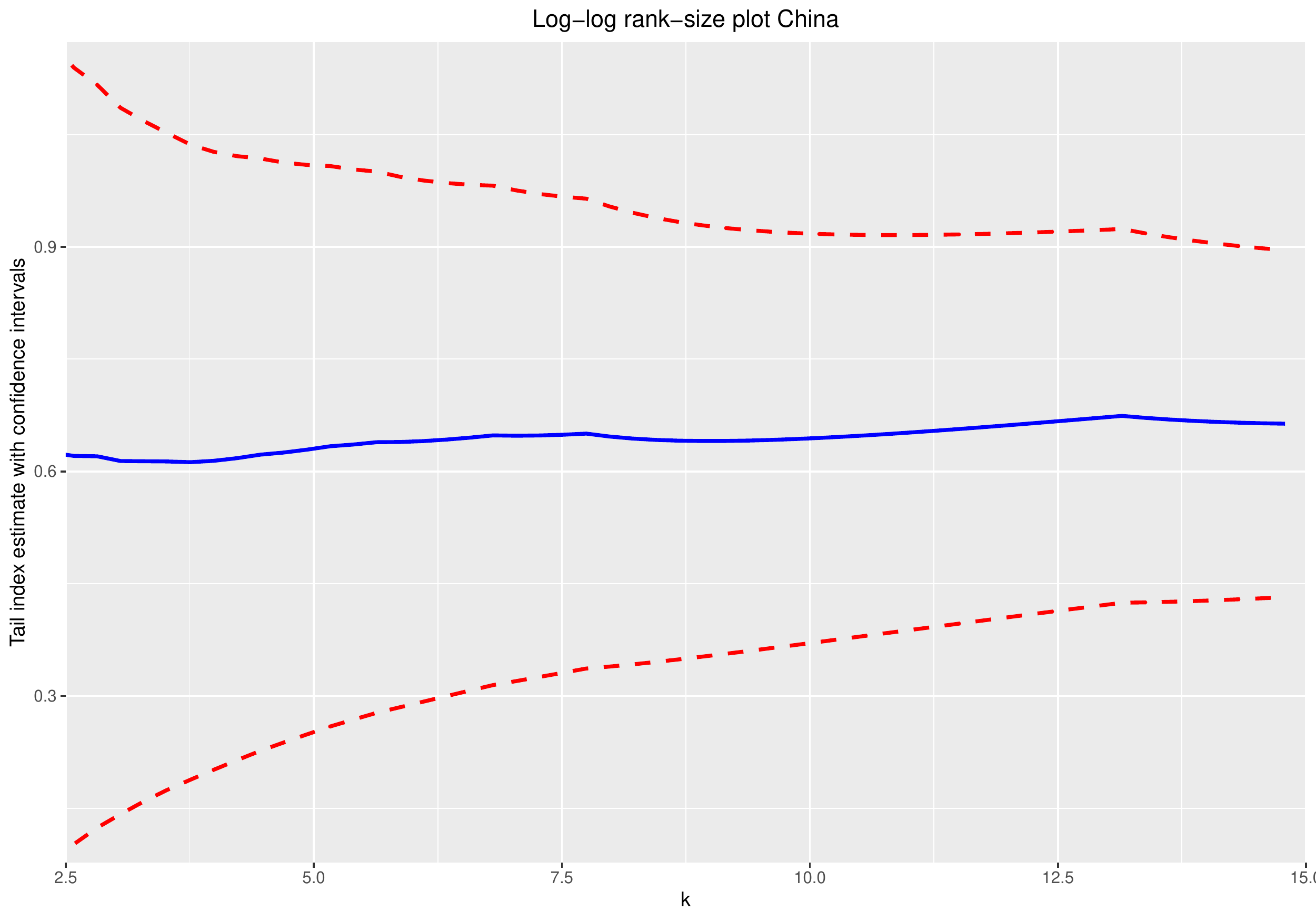}}\\
\subfigure[France]{\includegraphics[width=0.4\linewidth]{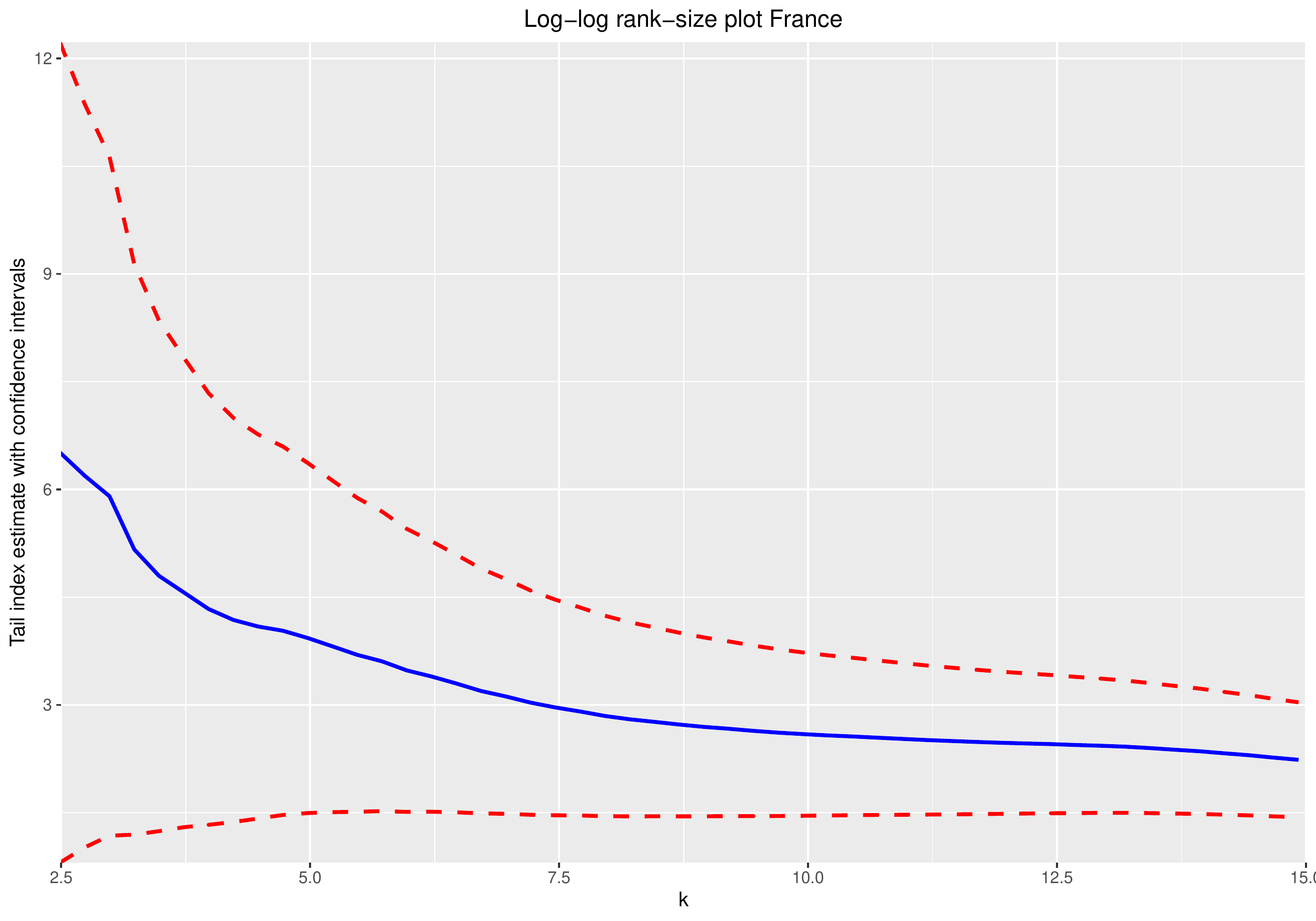}}
\subfigure[India]{\includegraphics[width=0.4\linewidth]{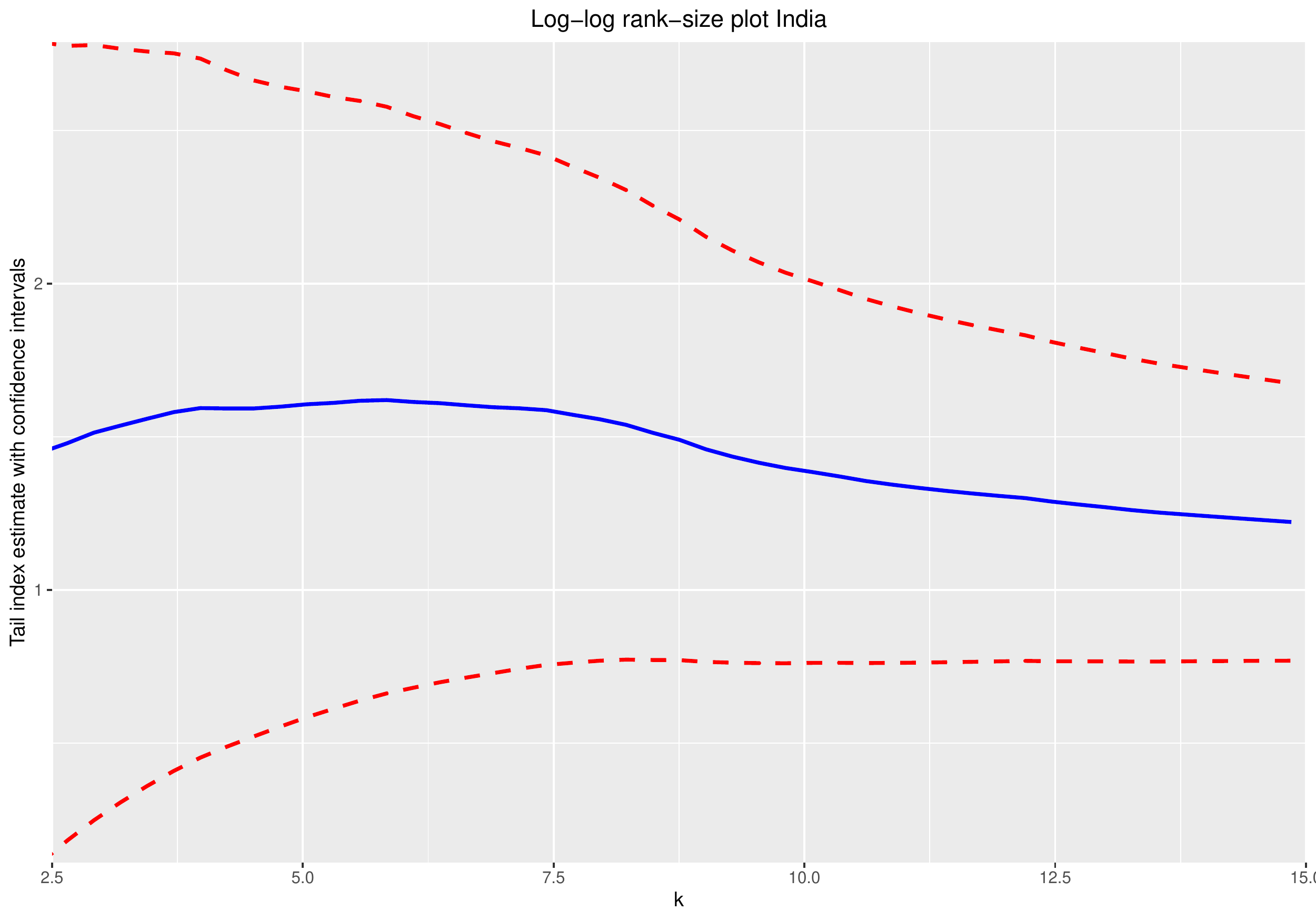}}\\
\end{center}%
\end{figure}

\setcounter{figure}{3}
\begin{figure}[h]%
\begin{center}%
\captcont{Log-log rank-size regression tail index estimates for positive changes in daily COVID-19 deaths (ctd)}
\subfigure[Italy]{\includegraphics[width=0.4\linewidth]{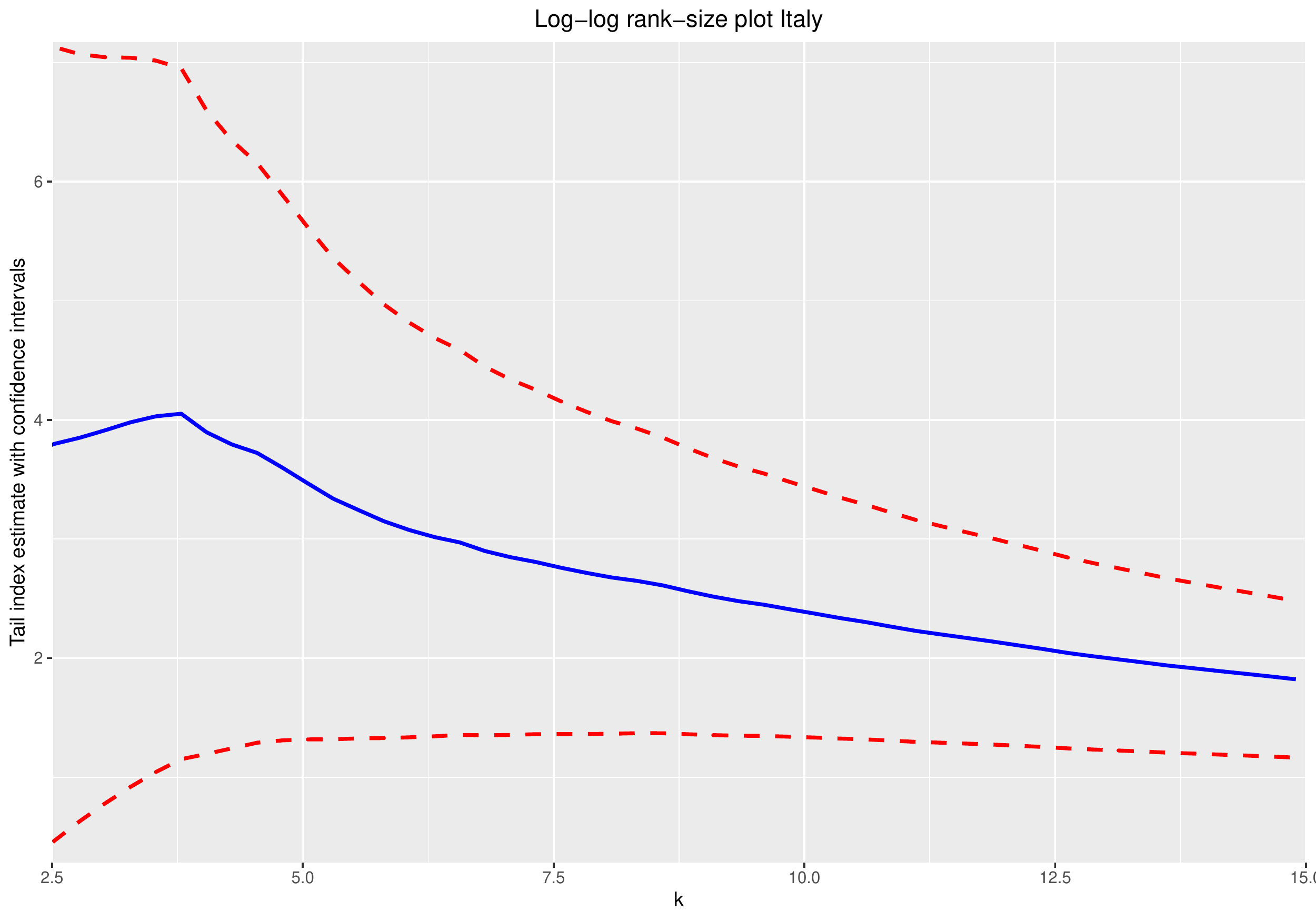}}
\subfigure[Russia]{\includegraphics[width=0.4\linewidth]{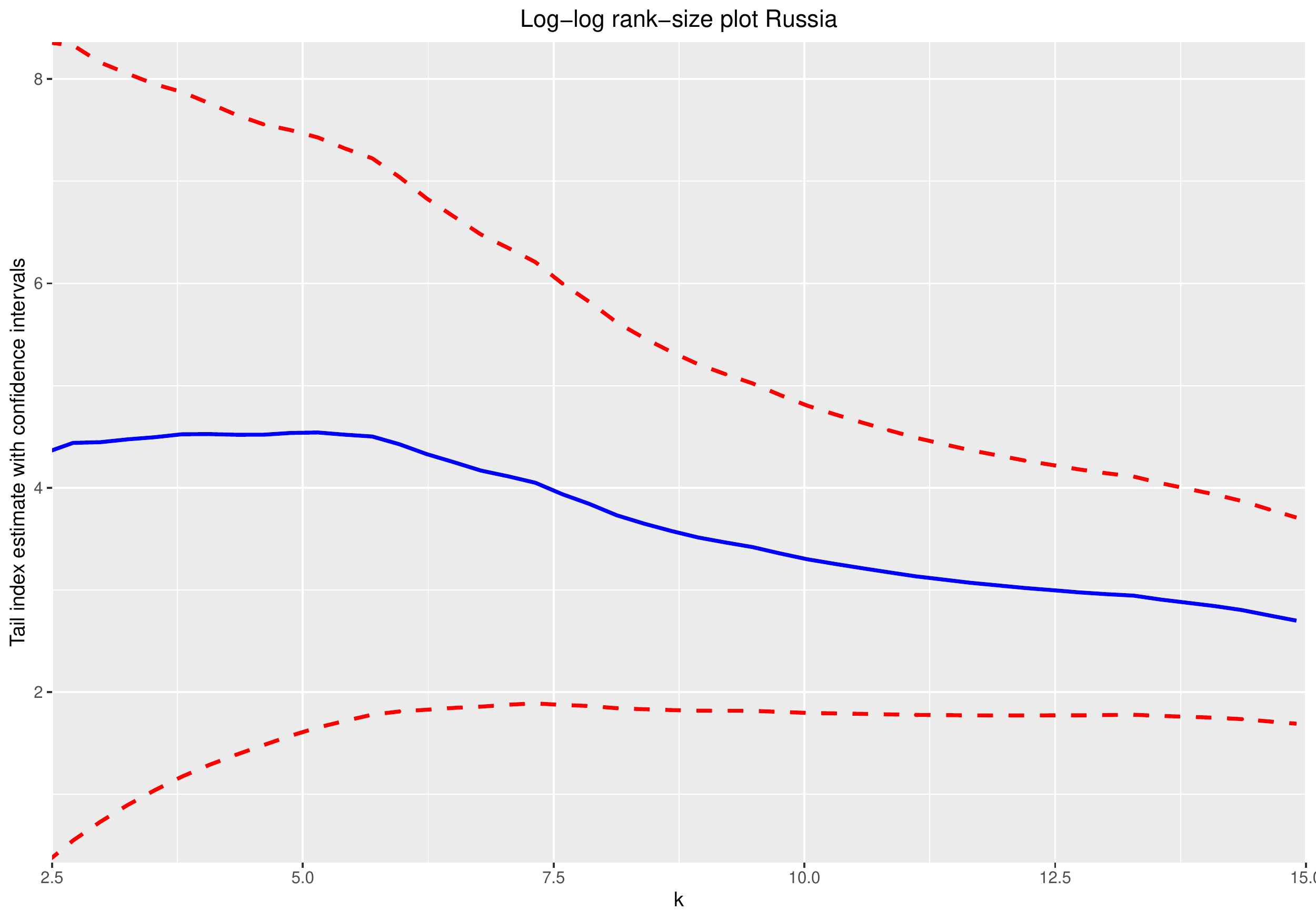}}
\subfigure[UK]{\includegraphics[width=0.4\linewidth]{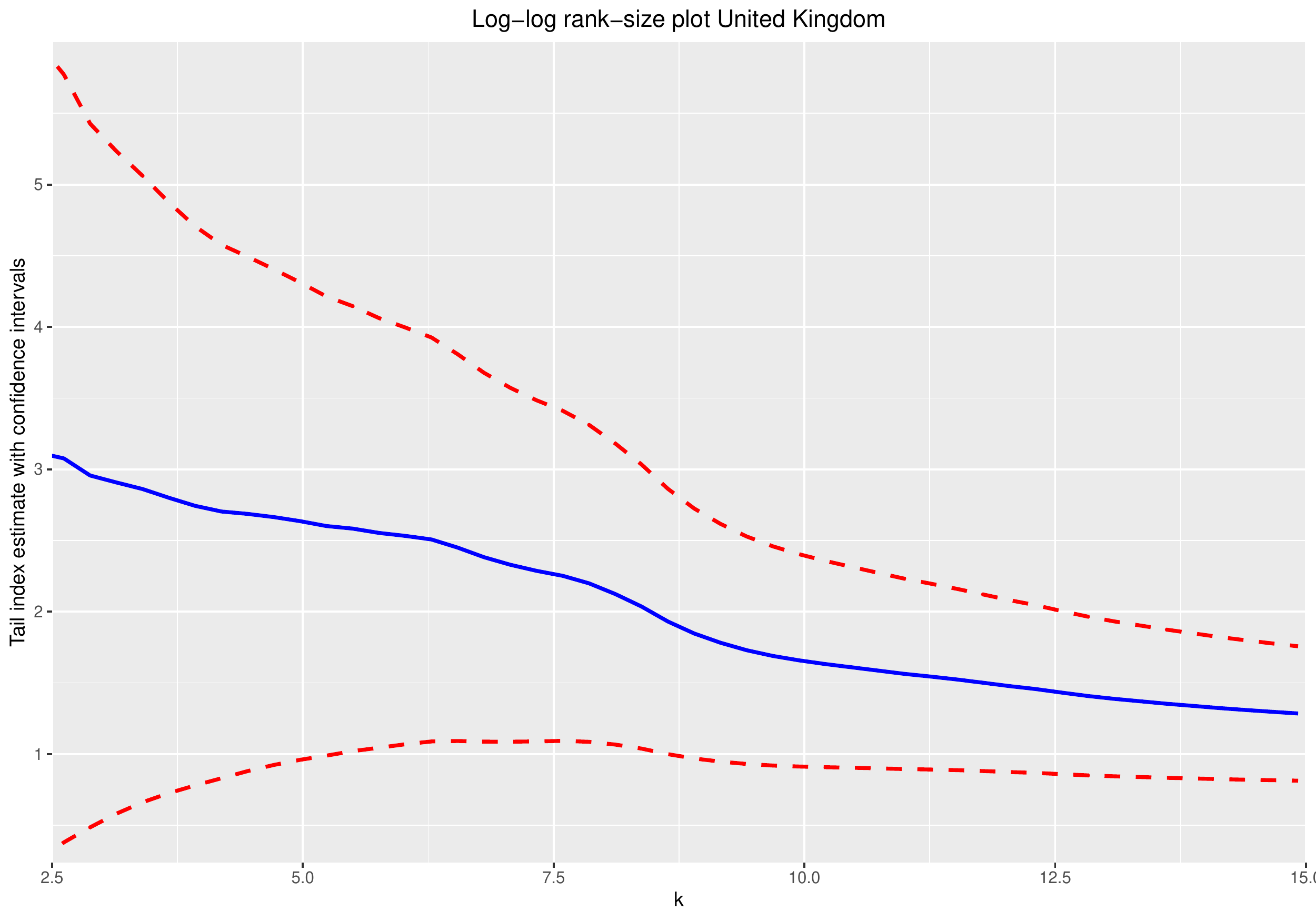}}
\subfigure[US]{\includegraphics[width=0.4\linewidth]{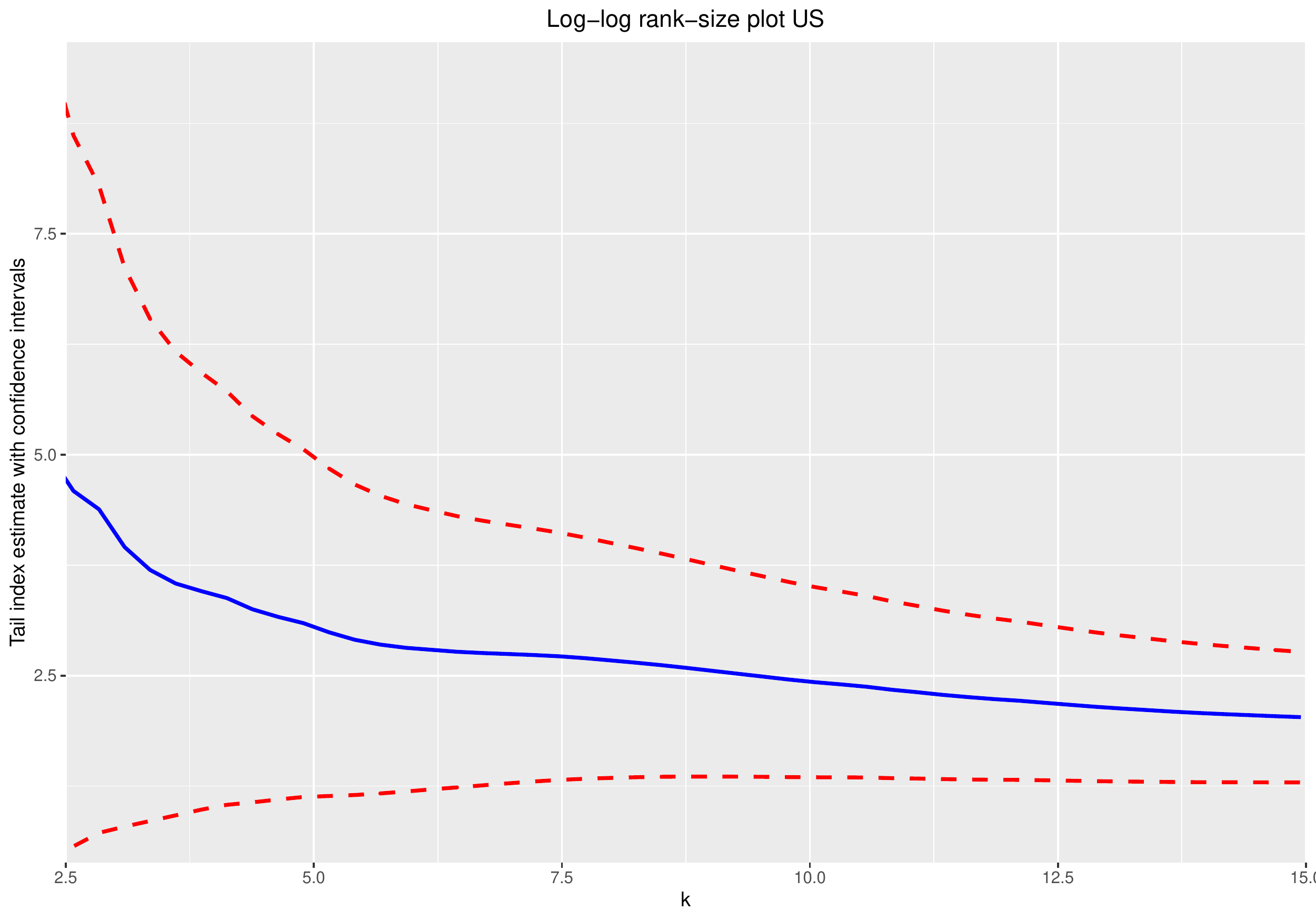}}
\end{center}
\end{figure}

\end{landscape}

\end{document}